\documentstyle[12pt]{article}

\catcode`\@=11
\long\def\@makefntext#1{
\protect\noindent \hbox to 3.2pt {\hskip-.9pt
$^{{\ninerm\@thefnmark}}$\hfil}#1\hfill}		

 \def\@makefnmark{\hbox to 0pt{$^{\@thefnmark}$\hss}}  

\def\ps@myheadings{\let\@mkboth\@gobbletwo
\def\@oddhead{\hbox{}
\rightmark\hfil\ninerm\thepage}
\def\@oddfoot{}\def\@evenhead{\ninerm\thepage\hfil
\leftmark\hbox{}}\def\@evenfoot{}
\def\sectionmark##1{}\def\subsectionmark##1{}}

\newcounter{sectionc}\newcounter{subsectionc}\newcounter{subsubsectionc}
\renewcommand{\section}[1] {\vspace{0.6cm}\addtocounter{sectionc}{1}
\setcounter{subsectionc}{0}\setcounter{subsubsectionc}{0}\noindent
	{\bf\thesectionc. #1}\par\vspace{0.4cm}}
\renewcommand{\subsection}[1] {\vspace{0.6cm}\addtocounter{subsectionc}{1}
	\setcounter{subsubsectionc}{0}\noindent
	{\it\thesectionc.\thesubsectionc. #1}\par\vspace{0.4cm}}
\renewcommand{\subsubsection}[1]
{\vspace{0.6cm}\addtocounter{subsubsectionc}{1}
	\noindent {\rm\thesectionc.\thesubsectionc.\thesubsubsectionc.
	#1}\par\vspace{0.4cm}}

\newcounter{appendixc}
\newcounter{subappendixc}[appendixc]
\newcounter{subsubappendixc}[subappendixc]

\renewcommand{\appendix}[1] {\vspace{0.6cm}
        \refstepcounter{appendixc}
        \setcounter{figure}{0}
        \setcounter{table}{0}
        \setcounter{equation}{0}
        \renewcommand{\thefigure}{\Alph{appendixc}.\arabic{figure}}
        \renewcommand{\thetable}{\Alph{appendixc}.\arabic{table}}
        \renewcommand{\theappendixc}{\Alph{appendixc}}
        \renewcommand{\theequation}{\Alph{appendixc}.\arabic{equation}}
        \noindent{\bf Appendix \theappendixc #1}\par\vspace{0.4cm}}

\def\abstracts#1{{
	\centering{\begin{minipage}{30pc}\tenrm\baselineskip=12pt\noindent
	\centerline{\tenrm ABSTRACT}\vspace{0.3cm}
	\parindent=0pt #1
	\end{minipage}}\par}}

\renewenvironment{thebibliography}[1]
	{\begin{list}{\arabic{enumi}.}
	{\usecounter{enumi}\setlength{\parsep}{0pt}
\setlength{\leftmargin 1.25cm}{\rightmargin 0pt}
	 \setlength{\itemsep}{0pt} \settowidth
	{\labelwidth}{#1.}\sloppy}}{\end{list}}

\newcounter{itemlistc}
\newcounter{romanlistc}
\newcounter{alphlistc}
\newcounter{arabiclistc}

\newcommand{\fcaption}[1]{
        \refstepcounter{figure}
        \setbox\@tempboxa = \hbox{\tenrm Fig.~\thefigure. #1}
        \ifdim \wd\@tempboxa > 6in
           {\begin{center}
        \parbox{6in}{\tenrm\baselineskip=12pt Fig.~\thefigure. #1}
            \end{center}}
        \else
             {\begin{center}
             {\tenrm Fig.~\thefigure. #1}
              \end{center}}
        \fi}

\newcommand{\tcaption}[1]{
        \refstepcounter{table}
        \setbox\@tempboxa = \hbox{\tenrm Table~\thetable. #1}
        \ifdim \wd\@tempboxa > 6in
           {\begin{center}
        \parbox{6in}{\tenrm\baselineskip=12pt Table~\thetable. #1}
            \end{center}}
        \else
             {\begin{center}
             {\tenrm Table~\thetable. #1}
              \end{center}}
        \fi}

\def\@citex[#1]#2{\if@filesw\immediate\write\@auxout
	{\string\citation{#2}}\fi
\def\@citea{}\@cite{\@for\@citeb:=#2\do
	{\@citea\def\@citea{,}\@ifundefined
	{b@\@citeb}{{\bf ?}\@warning
	{Citation `\@citeb' on page \thepage \space undefined}}
	{\csname b@\@citeb\endcsname}}}{#1}}

\newif\if@cghi
\def\cite{\@cghitrue\@ifnextchar [{\@tempswatrue
	\@citex}{\@tempswafalse\@citex[]}}
\def\citelow{\@cghifalse\@ifnextchar [{\@tempswatrue
	\@citex}{\@tempswafalse\@citex[]}}
\def\@cite#1#2{{$\null^{#1}$\if@tempswa\typeout
	{IJCGA warning: optional citation argument
	ignored: `#2'} \fi}}

\def\fnt#1#2{\footnotetext{\kern-.3em
	{$^{\mbox{\sevenrm #1}}$}{#2}}}

 1
 1
 1

\font\tenbf=cmbx10
\font\tenrm=cmr10
\font\tenit=cmti10

\font\ninerm=cmr9

\begin{document}
\pagestyle{empty}

\hspace{3cm}

\vspace{-2.5cm}
\rightline{{ FTUAM 95/20}}

\vspace{2.0cm}

\centerline{\tenbf SOFT SUPERSYMMETRY--BREAKING TERMS}
\baselineskip=16pt
\centerline{\tenbf AND THE $\mu$ PROBLEM\footnote{Based on talks given
at the Warsaw--Boston Workshop on "Physics from Planck scale to Electroweak
scale", Warsaw, September 1994; Korea Physical Society Fall Meeting,
Kyung--ju, October 1994; International Conference "Beyond the
Standard Model IV", Lake
Tahoe (California), December 1994; International Workshop on "Elementary
Particle Physics Present and Future", Valencia, June 1995.}}
\vspace{2.0cm}
\centerline{\tenrm C. MU\~NOZ\footnote{Research supported in part by:
the CICYT, under contract AEN93-0673; the European Union, under contracts
CHRX-CT93-0132 and SC1-CT92-0792.}}
\baselineskip=13pt
\centerline{\tenit Departamento de F\'{\i}sica Te\'{o}rica C-XI, Universidad
Aut\'{o}noma de Madrid}
\baselineskip=12pt
\centerline{\tenit Cantoblanco, 28049 Madrid, Spain}
\baselineskip=12pt
\centerline{\tenit cmunoz@ccuam3.sdi.uam.es}
\vspace{3.0cm}

\abstracts{The connection between Supergravity and the low-energy
world is analyzed. In particular, the soft Supersymmetry-breaking terms
arising in Supergravity, the $\mu$ problem and various solutions proposed
to solve it are reviewed. The soft terms arising in Supergravity
theories coming from Superstring theory are also computed and the solutions
proposed to solve the $\mu$ problem, which are naturally present in
Superstrings, are also discussed. The $B$ soft terms
associated are given for the different solutions. Finally, the low-energy
Supersymmetric-spectra, which are very charateristic, are obtained.}

\vspace{4.0cm}
\begin{flushleft}
{FTUAM 95/20} \\
{July 1995}
\end{flushleft}

\vfil
\rm\baselineskip=14pt
\newpage
\section{Introduction and Summary}
\noindent The  particle spectrum in Supersymmetric versions of the
Standard Model (SM) is in general determined by soft
Supersymmetry (SUSY)-breaking mass parameters like gaugino, squark, and slepton
masses. The possible numerical values of these soft terms are
only constrained by experimental bounds on SUSY-masses or
some indirect theoretical arguments.
If low-energy Supersymmetry is correct, eventually (and hopefully)
the spectrum of SUSY-particles will be measured and the structure of
soft SUSY-breaking terms will be subject to experimental test.
Since the breaking of Supergravity (SUGRA) generates the soft terms one could,
in principle, calculate explicitly their values. However, the functions which
determine the SUGRA theory are somehow arbitrary and therefore the soft terms
become SUGRA model-dependent. On the other hand, SUGRA theories coming from
Superstring theory are much more constrained and this will provide a theory
of soft terms which could enable us to interpret the (future) experimental
results on SUSY-spectra. The purpose of this section is to review this
approach for addressing the problem. After reviewing the theoretical
considerations which lead to the introduction of SUSY, the soft SUSY-breaking
terms are briefly analyzed in connection with SUGRA. Finally, after discussing
the relevant theoretical arguments that can be given in favour of Superstring
theory, the connection of this with SUGRA and therefore with the soft terms
is expounded.

\subsection{Why Supersymmetry?}

\noindent Despite the absence of experimental verification, relevant
theoretical
arguments can be given in favour of SUSY\cite{Rusos,Nilles}:

\vspace{0.2cm}

i) It is a new symmetry which relates bosons and fermions.

\noindent The importance of this is
twofold: we know from the past that symmetries are crucial in particle physics
and, besides, SUSY implies a new kind of unification,
between particles of different
spin. The latter involves that the Higgs is no longer a mysterious particle
as it stands in the SM, the only fundamental scalar particle
which should exist. Now, the Supersymmetric Standard Model (SSM) is naturally
full of fundamental scalars (squarks, sleptons and Higgses) related through
SUSY with their fermionic partners (quarks, leptons and Higgsinos).

\vspace{0.2cm}

ii) The local version of SUSY leads to a partial unification of the SM with
gravity, SUGRA\cite{Zumino,Nilles,history}.

\vspace{0.2cm}

iii) String theory\cite{SW,Green,history2} needs to be SUSY (Superstring)
in order to avoid tachyons in the spectrum and, besides, after compactification
of extra dimensions leads to an effective SUGRA.

\vspace{0.2cm}

iv) Whereas the quadratic and quartic terms of
the Higgs potential which are necessary in order to break the electroweak
symmetry have to be postulated "ad hoc" in the case of the SM, they appear in a
natural way in the context of the SSM.

\vspace{0.2cm}

v) The joining of the three gauge coupling constants of
the SM at a single unification scale which only taking into account SUSY
agrees with the experimental results.

\vspace{0.2cm}

\noindent But still, the most important argument
in favour of SUSY, is that

\vspace{0.2cm}

vi) It solves the so-called gauge hierarchy problem.

\noindent If we believe that the SM should be embedded within a more
fundamental theory
including gravity with a characteristic scale of order the Planck mass,
$M_P \equiv G_N^{-1/2}/\sqrt{8\pi}\simeq 2.4 \times 10^{18} GeV$,
then we are faced with the hierarchy problem. There is no symmetry protecting
the masses of the scalar particles against quadratic divergences in
perturbation
theory. Therefore they will be proportional to the huge cut-off scale
$\sim M_P$. The Higgs particle is included in the SM because of its good
properties: can have vacuum expectation value (VEV), without breaking
Lorentz invariance, inducing the spontaneous breaking of the electroweak
symmetry at the same time that generates the fermion masses through Yukawa
couplings. But, of course, all these properties are due to the fact that
the Higgs is a scalar particle. As mentioned above, this leads to a huge
mass for it and as a consequence for the $W$ and $Z$ gauge bosons.
This problem of naturalness, to stabilize $M_W \ll M_P$ against quantum
corrections,
is solved in SUSY since now the scalar mass and the mass of its superpartner,
the fermion, are related. As a consequence, only a logarithmic divergence
in the scalar mass is left. In diagrammatic language, the dangerous diagrams
of SM particles are cancelled with new ones which are present due to the
existence of the additional partners and couplings.

\subsection{Why soft Supersymmetry-breaking terms?}

\noindent SUSY cannot be an exact symmetry of nature and should be broken
since a SUSY particle with the same mass than its SM partner has never been
detected (e.g. there is no selectron with mass $\sim 0.5 MeV$). One may
introduce terms in the Lagrangian which explicitly break SUSY but, they
should not induce quadratic divergences in order not to spoil the
SUSY solution to the gauge hierarchy problem. This restricted type of terms
exists, they have been completely listed\cite{Grisaru},
and they are called soft SUSY-breaking terms. The simplest SSM is the
so-called Minimal Supersymmetric Standard Model (MSSM), where the matter
consists of three generations of quark and lepton superfields plus two Higgs
doublets (SUSY demands the presence of two Higgs doublets unlike
the SM where only one is needed),
$H_1$ and $H_2$ of opposite hypercharge,
and the gauge sector
consists of $SU(3)\times SU(2)_L\times U(1)_Y$ vector superfields. Assuming
certain universality of soft terms, these can be parametrized by only
four parameters: a universal gaugino mass (where the gauginos are the
superpartners of the
gauge bosons) $M$, a universal scalar mass $m$, a
universal trilinear
scalar parameter (associated with the Yukawa couplings) $A$ and an
extra bilinear scalar parameter (associated with possible bilinear
couplings) $B$.

The soft terms are very important since they determine the SUSY-spectrum,
like gaugino, squark and
slepton masses, and contribute to the Higgs potential generating the
radiative breakdown of the electroweak symmetry. Although in principle they
are constrained because they must reproduce the experimental results
($M_W \simeq 80 GeV$), the correct numbers can be obtained for wide ranges
of the above parameters. In this sense the predictivity of the theory is
limited.

\subsection{Supergravity as the origin of the soft terms}

\noindent The previous mechanism
for breaking SUSY explicitly looks arbitrary but it turns out to be
the most natural one when global SUSY is promoted to local, i.e. in SUGRA.
When SUGRA is spontaneously broken in a "hidden" sector
the soft SUSY-breaking terms are generated
(see e.g. refs.\cite{Hall,Soni} and references therein).
The process is the following: singlet scalar fields under
the observable gauge group ("hidden" sector fields) acquire VEVs giving rise
to spontaneous breaking of SUGRA. The goldstino, which is a combination of
the fermionic partners of the above fields, is swallowed by the gravitino
(the spin $3/2$ superpartner of the graviton in $N=1$ SUGRA)
which becomes massive. This
is the so-called super-Higgs effect. It is completely analogous to the usual
Higgs mechanism. Then, the hidden fields, which only have gravitational
interactions with the observable sector, decouple from the low-energy
theory and the only signals they produce are the soft terms. These are
characterized by the gravitino mass ($m_{3/2}$) scale and therefore in order
not to introduce a new problem of naturalness, $m_{3/2}$ should be of the
electroweak scale order (recall that the soft terms contribute to the
Higgs masses). An interesting non-perturbative source of
SUSY-breaking, capable of generating
this large mass hierarchy ($m_{3/2} \ll M_P$), is gaugino condensation in some
hidden sector
gauge group\cite{GC,Nilles2}.
Below $M_P$ (i.e. in the so-called flat limit where
$M_P \rightarrow \infty$ but $m_{3/2}$ is kept fixed),
one is left with a global supersymmetric
Lagrangian plus the soft SUSY-breaking terms. In summary, the MSSM is
just an effective
low-energy theory derived from SUGRA when this is spontaneously broken.

Two main arguments can be used in order to criticize the previous
SUSY-breaking mechanism: First of all, SUGRA is a non-renormalizable theory
and then it is not clear that we are doing a consistent analysis. Notice
however that the effective Lagrangian below the Planck scale is renormalizable
and we are interested only in this region. The general idea below this
approach is that we are
considering the SUGRA Lagrangian as an effective phenomenological Lagrangian
which comes from a bigger structure, renormalizable or even finite
(Superstring theory?). The situation is similar to the one of the old Fermi
Theory.
Second, the existence of the hidden sector has to be postulated "ad hoc".
However, we will see below that in the context of Superstring theory it appears
in
a natural way.

The full SUGRA Lagrangian\cite{Sugra} is specified in terms of two functions
which depend
on the hidden and observable scalars of the theory:
the real gauge-invariant K\"ahler function $G$ which is a combination
of two functions $K$ and $W$, and the analytic gauge
kinetic function $f$. $K$ is the K\"ahler potential whose second
derivative determines the kinetic
terms for the fields in the chiral supermultiplets and is thus important
for obtaining the proper normalization of the fields. $W$ is the complete
analytic
superpotential which is related with the Yukawa couplings (which eventually
determine the fermion masses) and also includes possibly non-perturbative
effects. Finally, $f$ determines the kinetic terms for
the fields in the vector
supermultiplets, and in particular the gauge coupling constant
$Re f_a=1/g_a^2$. The subindex $a$ is associated with the different
gauge groups of the theory $G=\prod_a G_a$.
For example, in the case of the SM coupled to SUGRA it is associated
with $SU(3)$, $SU(2)_L$, $U(1)_Y$.
Then, once we know these functions the soft SUSY-breaking terms are
calculable. Unfortunately for the predictivity of the theory, $G$ and $f$ are
arbitrary and therefore
{\it the soft terms become SUGRA model-dependent}.
All the above mentioned problems can be solved in Superstring theory.

\subsection{Why Superstring theory?}

\noindent As in the case of SUSY, relevant theoretical arguments can be given
in favour
of Superstring theory:

\vspace{0.2cm}

i) It is possible to obtain models resembling the SSM at
low-energy\cite{Modelos,Search}.

\noindent Of
course, this is crucial in order to connect Superstring theory with the
observable
world.

\vspace{0.2cm}

ii) It is the only hope to answer fundamental questions that in the
context of the SM, SSM or Grand Unified Theories (GUTs) cannot even be posed:
why the gauge and Yukawa couplings should have a particular value?.

\noindent First,
the gauge coupling constants are dynamical because they arise as the VEV of
a gauge singlet field $S$ called the dilaton, $\langle Re S\rangle = 1/g_a^2$.
This result can be understood taking into account that in
Superstring theory $f_a \simeq S$ (at tree level\cite{Witten}). It is worth
noticing here that
the gauge coupling constants
are unified\footnote{The natural
unification scale in Superstring models is\cite{Kaplunovsky}
$M_{string}\simeq 0.5 \times g_{string}
\times  10^{18}$ GeV, where $g_{string}=(ReS)^{-1/2}\simeq 0.7$. However,
the (apparent success of the)
joining of gauge coupling constants at high energies,
with the particle content of the
MSSM, takes place at a scale $M_X\simeq 3\times 10^{16}$ GeV.
Although this unification takes place at energies around a factor $\simeq 12$
smaller than expected in Superstring theory, several mechanisms have been
proposed in order to explain this discrepancy\cite{Yo}.} even in the
absence of a GUT. Thus GUT gauge groups, as e.g.
$SU(5)$ or $SO(10)$, are not mandatory in order to have unification in the
context of Superstring theory. Second, the Yukawa couplings, which determine
the
quark and lepton masses, can be explicitly calculated\cite{Hamidi,Faustino} and
they turn out to be
also dynamical. They depend in general on other gauge singlet fields $T_m$
called the moduli whose VEVs
determine the size and shape of the
compactified space. E.g. for the overall
modulus $\langle Re T\rangle \sim R^2$.
In fact, particular values of these fields allow us to reproduce,
in principle, the peculiar
observed pattern of quark and lepton masses and mixing angles\cite{Faustino}.
Since
general experimental data (values of the gauge couplings and no observation
of extra dimensions) demand
$\langle Re S\rangle \sim 2$ and
$\langle Re T\rangle \sim 1$ (in $M_P$ units), the initial
questions translate as how are the VEVs determined. This will be discussed
below.

\vspace{0.2cm}

\noindent Finally, the most outstanding virtue of Superstring theory is that

\vspace{0.2cm}

iii) It is the only (finite) theory
which can unify all the known
interactions including gravity.

\subsection{Superstring theory as the origin of Supergravity}

\noindent The ten-dimensional Heterotic Superstring\cite{Gross}
after compactification of six extra dimensions (on some compact manifolds)
leads to a $N=1$ effective SUGRA. Now, $f$ and $K$ {\it are in principle
calculable}\cite{Witten,FKP,Kaplunovsky,DKL1,DKL2,Louis}
from Superstring scattering
amplitudes. Besides, whereas in SUGRA
(non-Superstring) models we do not have the slightest idea
of what fields could be involved in SUSY-breaking, four-dimensional
{\it Superstring theory automatically has natural candidates for that job}: the
dilaton $S$ and the moduli $T_m$. These gauge singlet fields are
generically present in four-dimensional Heterotic Superstrings since $S$
is related with the gravitational sector of the theory and $T_m$ are related
with the extra dimensions. While other extra fields could also play
a role in specific models, the dilaton and moduli constitute in some way
the {\it minimal possible SUSY-breaking sector in Superstring theory}\footnote
{Starting with this minimal sector one can also study the possible role
on SUSY-breaking of other extra fields (see the example discussed in
section 8 of ref.\cite{BIM}).}.

Concerning the superpotential $W$, the situation
is more involved. It is known that the process of SUSY-breaking in
Superstring theory has to have a non-perturbative origin
since SUSY is preserved order by order in perturbation
theory and hence $S$ and $T_m$ are
undetermined at this level (the scalar potential $V(S,T_m)$ is flat).
On the other hand, very little is known about non-perturbative
effects in Superstring theory, particularly in the four-dimensional case.
It is true
that gaugino condensation gives rise to an effective $W(S,T_m)$ which breaks
SUSY
at the same time that $S$ and $T_m$ acquire reasonable VEVs, as the
ones explained in point (ii) (see e.g. ref.\cite{Beatriz} and references
therein), and determines
explicitly the values of the soft SUSY-breaking terms\cite{Beatriz2}. However
one should
keep in mind two caveats. First, the VEV of the scalar potential, i.e. the
cosmological constant (for a review of the cosmological constant
problem, see ref.\cite{cc}), is
non-vanishing
(negative). For an
extended discussion on this point see section 8 of ref.\cite{BIM}. Second,
this analysis requires the assumption that the dominant
non-perturbative effects in Superstring theory are the field theory ones. This
is
because gaugino condensation is not a pure "stringy" mechanism.
Thus a pessimist would say that Superstring theory does not look particularly
promising in trying to get information about the SUSY-breaking sector
of the theory.

However, since $K$ and $f$ are known in Superstring models, and the degrees of
freedom involved in the process of SUSY-breaking have been identified
(the hidden fields $S$ and $T_m$), the effect of SUSY-breaking can be
parametrized by the VEVs of the auxiliary fields of the degrees of freedom
identified without specifying what is the origin of SUSY-breaking.
This situation is similar to the one concerning
$SU(2)_L\times U(1)_Y$ breaking in the SM.
We do not really know for sure how the gauge symmetry of the SM is
broken. We just parametrize our ignorance by using a Higgs field
(either composite or elementary) with a non-vanishing VEV,
the key ingredient here is knowing the
degrees of freedom (an $SU(2)_L$ doublet) involved in the process of
symmetry breaking. Once we know that, we can obtain all the
experimentally confirmed predictions of the SM.
The purpose of the present paper
is to review the above approach for addressing the problem, trying to provide
a theory of soft terms which could enable us to interpret the (future)
experimental results on SUSY-spectra.

The structure of the paper is as follows. In sect.2 we discuss some general
formulae for the computation of soft terms in SUGRA. We analyze how the
existence of hidden sector dependent masses and couplings modifies the usual
soft terms taking a simple case as a guiding example. Finally, we also
discuss the so-called $\mu$ problem and various solutions
proposed to generate a $\mu$ term in the low-energy theory. The $B$ soft term
is given for the different solutions. In sect.3 we apply
the formulae obtained
in the previous section to the computation of soft terms in
Superstring
theory\cite{Gaugino,Cvetic,IL,Beatriz,Beatriz2,KapluLouis,BIM,Ferrara}.
It
turns out to be specially useful to introduce a "goldstino angle" whose
value tells us where the dominant source of SUSY-breaking
resides\cite{BIM}. All
formulae for soft parameters take on particularly simple forms when written
in terms of this variable. We also allow for a non-vanishing vacuum energy
and arbitrary complex phases in the relevant VEVs. Allowing for these turns
out to be important for some relevant issues concerning soft terms. We also
discuss the more model-dependent $B$ soft term for the solutions to
the $\mu$ problem which were analyzed in the previous section. This kind of
solutions are naturally present in Superstring theory.
Finally, we compute the soft terms for some large classes of models
including the large-radius limit of
Calabi--Yau-type-compactifications\cite{Horowitz}
and orbifold-type models\cite{Harvey}.
The low-energy renormalization group running of soft terms, the
low-energy sparticle-spectra and the appropriate radiative
$SU(2)_L\times U(1)_Y$ breaking are considered.
This analysis leads to
specific patterns for the SUSY-spectra which could be tested
in future colliders.

\section{Soft terms from Supergravity and the $\mu$ problem}
\vspace{-0.7cm}
\subsection{General structure of soft terms}

\noindent Soft scalar masses, trilinear and bilinear scalar
terms arise from the expansion of
the SUGRA
scalar potential
\begin{eqnarray}
V=e^{G} \; [ \; G_{\alpha} \; (G^{-1})^{\alpha}_{\beta} \; G^{\beta}-3 \; ]
\label{F1}
\end{eqnarray}
and soft gaugino masses for the canonically normalized gaugino fields
can be obtained from the fermionic part of the SUGRA Lagrangian\cite{Sugra}
\begin{eqnarray}
M_a=\frac{1}{2} \;{(Ref_a)}^{-1}\; e^{G/2} \; f_a^\alpha \;
(G^{-1})_\alpha^\beta\ G_\beta\ ,
\label{F2}
\end{eqnarray}
where the real gauge-invariant K\"ahler function $G$ is given by
\begin{eqnarray}
G(z_\alpha,z_\alpha^{*})=K(z_\alpha,z_\alpha^{*})+\log|W(z_\alpha)|^2
\label{G}
\end{eqnarray}
and we use from now on the standard SUGRA mass units where $M_P=1$ and
the standard SUGRA conventions on
derivatives (e.g. $G_\alpha = \frac{\partial G}{\partial z_\alpha^{*}}$,
$G^\alpha = \frac{\partial G}{\partial z_\alpha}$).
$K$  and $W$ are given in general by the form
\begin{eqnarray}
K &=& K_0(h_l,h_l^*) + K_{ij}\phi_i\phi_j^* + (Z_{ij}\phi_i\phi_j + h.c.)
     +... \ ,
\label{F3}\\
W &=& W_0(h_l) +\mu_{ij}\phi_i\phi_j + Y_{ijk}\phi_i\phi_j\phi_k +... \ ,
\label{F4}
\end{eqnarray}
where we assume two different types of scalar fields
$z_\alpha=h_l,\phi_i$. $\phi_i$ correspond to the observable sector
(they include the SSM fields) and $h_l$ correspond to a hidden sector.
The latter are responsible for the SUSY breaking when some of them
acquire large ($\gg M_W$) VEVs.
The ellipsis indicates terms of higher order in $\phi_i,\phi_i^*$ and {\it the
quantities $\mu_{ij}, Y_{ijk}, K_{ij}$ and $Z_{ij}$ are in general
$h_l, h_l^*$ dependent}.

Then, the form of the effective soft Lagrangian obtained from
eqs.(\ref{F1},\ref{F2}) is given in general by
\begin{eqnarray}
{\cal L}_{soft}= \frac{1}{2} \sum_a M_a \bar{\widehat{\lambda}}_a
  \widehat{\lambda}_a - \sum_i m_i^2 |\widehat{\phi}_i|^2 -
(A_{ijk} \widehat{Y}_{ijk}
	    \widehat{\phi}_i \widehat{\phi}_j \widehat{\phi}_k
  +B_{ij} {\widehat{\mu}}_{ij} \widehat{\phi}_i \widehat{\phi}_j+h.c.)\ .
\label{F6}
\end{eqnarray}
The mass parameter $\widehat{\mu}$ is related with $Z$ and
$\mu$ terms of eqs.(\ref{F3}) and (\ref{F4}) respectively and will be
discussed below in the context of the so-called $\mu$ problem.
We recall that the passage to the effective low-energy theory involves
a number of rescalings.
In particular, in the previous equation
\begin{eqnarray}
\widehat{\phi}_i    &=& K_i^{1/2} \phi_i\ ,
\nonumber\\
\widehat{\lambda}_a &=& (Re f_a)^{1/2} \lambda_a\ ,
\nonumber\\
{\widehat{Y}}_{ijk}    &=& Y_{ijk} \; \frac{W_0^*}{|W_0|} \; e^{K_0/2} \;
(K_i K_j K_k)^{-1/2}\ ,
\label{F7}
\end{eqnarray}
where $\widehat{\phi}_i, \widehat{\lambda}_a$ are the scalar and
gaugino canonically
normalized fields respectively\footnote{For
phenomenological reasons related to the
absence of flavour changing neutral currents (FCNC) in the effective
low-energy theory (see section 5 of ref.\cite{BIM} for a discussion on
this point) from now on we will assume a diagonal form for the part
of the K\"ahler potential associated with matter fields,
$K_{ij}=K_{i}\delta_j^i$ in eq.(\ref{F3}).}.

In the case of the MSSM the K\"ahler potential (to first order in the
observable fields $\phi_i$) and
superpotential have the form (see eqs.(\ref{F3},\ref{F4}))
\begin{eqnarray}
K &=& K_0(h_l,h_l^*)+ \sum_i K_i \phi_i \phi_i^* + ( Z H_1 H_2 + h.c. )\ ,
\label{F9}\\
W &=& W_0(h_l)+\mu H_1 H_2 + \!\! \sum_{generations} \!\!
     (Y_u Q_L H_2 u_L^c + Y_d Q_L H_1 d_L^c + Y_e L_L H_1 e_L^c)\ ,
\label{F10}
\end{eqnarray}
where now $ i=Q_L, u_L^c, d_L^c, L_L,e_L^c, H_1, H_2 $. These equations include
the usual Yukawa couplings ($ Y_{ijk}=Y_u,Y_d,Y_e $, in a
self-explanatory notation) and {\it we have allowed a possible
mass $\mu$ and coupling $Z$
for the Higgses} (recall that they have
opossite hypercharges), where $\mu$ and $Z$ are in principle free parameters.
Finally, the subindex $a$ in eq.(\ref{F2}) is associated with the different
gauge groups of the theory, i.e. $SU(3)$, $SU(2)_L$ and $U(1)_Y$.

As already explained in the Introduction, the particular values of the
soft terms depend on the type of SUGRA theory from which the MSSM derives
and, in general, on the mechanism of SUSY-breaking. But, in fact, is still
possible to learn things about soft terms without knowing the details
of SUSY-breaking. Let us consider the simple case of canonical kinetic
terms for hidden and observable fields\cite{Bfs,Hall} (i.e.
$K_0=\sum_l h_l h_l^*$ and $K_i=1$ in eq.(\ref{F9})) and $Z=0$.
Then, irrespective of the SUSY-breaking mechanism, the scalar masses and
the $A$, $B$ terms can be straightforwardly computed:
\begin{eqnarray}
m_i^2	&=& m_{3/2}^2 + V_0\ ,
\label{F12} \\
A_{ijk} &=& m_{3/2} \sum_l {G_0}_{h_l} \left( h_l^*
            + \frac{Y_{ijk}^{h_l}}{Y_{ijk}} \right)\ ,
\label{F13} \\
B_{\mu} &=& m_{3/2} \sum_l {G_0}_{h_l}
	    \left( h_l^* + \frac{\mu^{h_l}}{\mu}\right) - m_{3/2}\ ,
\label{F14}
\end{eqnarray}
where $m_{3/2}^2=e^{G_0}$ is the gravitino mass and
$V_0=e^{G_0}(\sum_l |{G_0}_{h_l}|^2 -3)$
is the VEV of the scalar potential (i.e. the cosmological
constant). The latter has a bearing on measurable quantities like scalar
masses and therefore the way we deal with the cosmological constant problem
is important. Anyway, the scalar masses are automatically universal in this
case. This also happens for the $A$ terms assuming that the Yukawa couplings,
$Y_{ijk}$, are
hidden field independent. Besides, the gaugino masses are universal
if the gauge kinetic function is the same
for the different gauge groups of the theory $f_a=f$
(note that e.g. the case of a
constant $f$ is not phenomenologically interesting since
it would imply $M_a=0$ as can be seen from eq.(\ref{F2})). Finally, a
particularly interesting value of $B$ can be obtained assuming also that
$\mu$ is $h_l$
independent and the cosmological constant is vanishing. Then, from
eqs.(\ref{F13},\ref{F14}) the well known result for the $B$ term is
recovered:
\begin{eqnarray}
B_{\mu} = A-m_{3/2} \ ,
\label{F15}
\end{eqnarray}
where $A=A_{ijk}$ and we call
$B_\mu$ the $B$ term since it is associated only (recall that
we are analyzing the case $Z=0$) with the $\mu$ term of eq.(\ref{F10}).

This SUGRA theory is attractive for its simplicity and for the natural
explanation that it offers to the universality of the soft scalar masses.
Actually, universality is a desirable property not only to reduce the
number of independent parameters in the MSSM, but also for phenomenological
reasons, particularly to avoid flavour-changing neutral currents (FCNC)
(see e.g. ref.\cite{Rossbook}).

However, One can think of many possible
SUGRA models (with different $K$ and $f$) leading to SUSY-breaking in a
hidden sector of the theory
{\it leading in turn to different results for the soft terms}. We will see
in sect.3 how this problem can be ameliorated in the context of Superstring
theory, where $K$, $f$ and the hidden sector are more constrained.

\subsection{The $\mu$ problem}

\noindent Let us now discuss the $\mu$ problem and the different
solutions proposed
in the literature illustrating them in the case of canonical kinetic
terms. From eqs.(\ref{F6},\ref{F12},\ref{F14}) and taking into
account SUSY D and F-terms the
relevant Higgs scalar potential along the neutral direction for the
electroweak breaking is readily obtained
\begin{eqnarray}
V(H_1,H_2)= \frac{1}{8} (g_2^2+{g'}^2) ( |H_2|^2 - |H_1|^2 )^2
	    + m_1^2 |H_1|^2 + m_2^2 |H_2|^2 - m_3^2 ( H_1 H_2 + h.c. )
\label{F16}
\end{eqnarray}
with
\begin{eqnarray}
m_{1,2}^2     &=& m_{3/2}^2 + V_0 + |\widehat{\mu}|^2\ ,
\nonumber\\
m_3^2	      &=& -B_\mu \widehat{\mu}\ ,
\nonumber\\
\widehat{\mu} &=& \mu \; e^{K_0/2} \; \frac{W_0^*}{|W_0|}\ ,
\label{F17}
\end{eqnarray}
where $g_3=g_2=g_1=\sqrt{5/3}\;g'$ at the unification scale and the
Higgsino mass
$\widehat{\mu}$ gives a SUSY contribution (through the F-terms) to the
Higgs scalar masses. This is the SUSY version of the usual Higgs
potential in the SM. As mentioned in the Introduction, it appears in a
natural way whereas in the SM has to be postulated "ad hoc".
It must develope a minimum at $\langle H_{1,2}\rangle = {\nu}_{1,2}$ in such a
way that ${\nu}_1^2+{\nu}_2^2=2M_W^2/g_2^2$. This is the realistic minimum that
corresponds to the standard vacuum.

For this scheme to work, the presence of the second term in
eq.(\ref{F10}) is crucial. If $\mu=0$, then the form of the
renormalization group equations (RGEs) implies that such a term
is not generated at any $Q$ scale since $\mu(Q)\propto \mu$.
The same occurs for $m_3$, i.e. $m_3(Q)\propto \mu$.
Then, the minimum of the potential of eq.(\ref{F16}) occurs
for ${\nu}_1=0$ and, therefore, $d$-type quarks
and $e$-type leptons remain massless. Besides, the superpotential
of eq.(\ref{F10}) with $\mu=0$ possesses a spontaneously broken Peccei-Quinn
symmetry\cite{Peccei} leading to the appearance of an unacceptable
Weinberg--Wilczek axion\cite{Weinberg}.

Once it is accepted that the presence of the $\mu$-term
in the superpotential is essential, there arises an inmediate
question: Is there any dynamical reason why $\mu$ should be
small, of the order of the electroweak scale?. Note that,
to this respect, the $\mu$-term is different from the soft
SUSY-breaking terms,
which are characterized by the small scale $m_{3/2}$ once
we assume correct SUSY breaking. In principle the natural scale
of $\mu$ would be $M_P$, but this would re-introduce the
problem of naturalness since the Higgs scalars get a contribution
$\mu^2$ to their squared mass (see eq.(\ref{F17})). Thus,
any complete explanation of the electroweak breaking scale
must justify the origin of $\mu$. This is the so-called
$\mu$ problem. This problem has been considered by
several authors and different possible solutions
have been proposed\cite{Hall,Kim,Masiero,Casas,Giudice,Ignatios}.

{\it (a)} In ref.\cite{Casas} was pointed out that the presence of a
non-renormalizable term in the superpotential
\begin{eqnarray}
\lambda W_0 H_1 H_2
\label{F18}
\end{eqnarray}
characterized by the coupling $\lambda$, which
mixes the observable sector with the hidden sector, yields dynamically a
$\mu$ parameter when $W_0$ acquires a VEV
\begin{eqnarray}
\mu = \lambda W_0\ .
\label{F19}
\end{eqnarray}
The fact that $\mu$ is of the electroweak scale order is a consequence
of our assumption of a correct SUSY-breaking scale
$ m_{3/2} = e^{K_0/2} |W_0| = O(M_W) $.
Now, with this solution to the $\mu$ problem, eq.(\ref{F14}) gives
(let us call $B_{\lambda}$ the $B$-term associated with
eqs.(\ref{F18},\ref{F19}))
\begin{eqnarray}
B_{\lambda} = m_{3/2} \left (2+ \frac{V_0}{m_{3/2}^2} +
	 \sum_l {G_0}_{h_l} \; \frac{\lambda^{h_l}}{\lambda} \right)
\label{F20}
\end{eqnarray}
and $\widehat{\mu} = m_{3/2} \lambda $ in eq.(\ref{F17}). For $\lambda$
independent of $h_l$ a very simple result is obtained
\begin{eqnarray}
B_{\lambda} = m_{3/2} \left (2+ \frac{V_0}{m_{3/2}^2} \right)\ .
\label{F21}
\end{eqnarray}
The value of $A$ is still given by eq.(\ref{F13}), but the relation
$B_{\lambda} = A-m$ (see eq.(\ref{F15})) is no longer true even in the case
of $Y_{ijk}$ independent of $h_l$. For this mechanism to work the $\mu H_1 H_2$
term in eq.(\ref{F10}) must be absent (otherwise the natural scale for $\mu$
would be the Planck mass), a fact that remarkably enough, is automatically
guaranteed in the framework of Superstring theory as we will see below.

{\it (b)} In refs.\cite{Masiero,Casas} was shown
that if the coupling $Z$ is present in the
K\"ahler potential (see eq.(\ref{F9})), an effective low-energy $B$-term
is naturally generated of the order of magnitude of the gravitino mass.
We will call it $B_Z$.
We can compute its form in our guiding example with
canonical kinetic terms.
The result is
%
\begin{eqnarray}
B_Z &=& \frac{m_{3/2}}{X} \left( 2 + \frac{V_0}{m_{3/2}^2}
       + \sum_l [{G_0}_{h_l} \frac{Z^{h_l}}{Z} - G_0^{h_l} \;
\frac{Z_{h_l}}{Z}
       - |{G_0}^{h_l}|^2 \; \frac{Z_{h_l}^{h_l}}{Z}] \right)\ ,
\nonumber\\
X   &\equiv& 1- \sum_l G_0^{h_l} \; \frac{Z_{h_l}}{Z}
\label{F22}
\end{eqnarray}
and now the Higgsino mass $\widehat{\mu}$ in eq.(\ref{F17}) is modified
to $ \widehat{\mu} = m_{3/2} X Z $. Notice that the dependence on
$Z$ and $\lambda$ in eqs.(\ref{F22},\ref{F20}) is the same for the case of
an analytic $Z$, i.e. $Z$ independent
of $h_l^*$.
This is not a surprising result since both mechanisms
for solving the $\mu$ problem are
equivalent in this particular case\cite{Casas}.
Indeed, in the case now considered, the SUGRA theory is
equivalent to the one with a K\"ahler potential $K$ (without the
terms $Z H_1 H_2 + h.c.$ of eq.(\ref{F9})) and a superpotential
$ W e^{Z H_1 H_2} $, since the function $ G = K + \log |W|^2 $ is
the same for both.
After expanding the exponential, the superpotential will have a
contribution $Z W_0 H_1 H_2$, i.e. a term of the type of eq.(\ref{F18}).
Of course, in the simple case $Z$ also independent of $h_l$, $B_Z$
coincides with the result of eq.(\ref{F21})
\begin{eqnarray}
B_Z = m_{3/2} \left (2+ \frac{V_0}{m_{3/2}^2} \right)\ .
\label{F23}
\end{eqnarray}
Besides, assuming vanishing cosmological constant, i.e. $V_0=0$, the value
of $B$ in both cases {\it (a)}, {\it (b)} is given by
\begin{eqnarray}
B = 2 m_{3/2} \ .
\label{F323}
\end{eqnarray}
%


Notice that it is conceivable that both mechanisms could be present
simultaneously.
In that case the general expressions for the $B$-term and Higgsino mass
are easily obtained
\begin{eqnarray}
B &=& \frac{1}{\widehat{\mu}}( B_{\lambda} m_{3/2} \lambda + B_Z m_{3/2} XZ )
\ ,
\nonumber\\
\widehat{\mu} &=& m_{3/2} \lambda + m_{3/2} X Z\ ,
\label{F24}
\end{eqnarray}
where $B_{\lambda}$ and $B_Z$, $X$ are given in eqs.(\ref{F20}) and (\ref{F22})
respectively.

{\it (c)} In refs.\cite{Hall,Giudice} the observation was made that in the
framework of any SUSY-GUT, starting again with $\mu=0$, an effective $\mu$
term is generated by the integration of the heavy degrees of freedom. The
prediction for $B$ is once more given by eq.(\ref{F323}).

The solutions discussed here in order to solve the $\mu$ problem
{\it are naturally present in Superstring theory\footnote{Another
possible solution via SUSY-breaking in Superstring perturbation theory
can be found in ref.\cite{Ignatios}.}}.
In ref.\cite{Casas} was first remarked that the $\mu H_1 H_2$ term
(see eq.(\ref{F10})) is naturally absent from $W$ (otherwise the
natural scale for $\mu$ would be $M_P$) since in SUGRA theories
coming from Superstring theory mass terms for light fields are forbidden
in the superpotential. Then a realistic example where non-perturbative
SUSY-breaking mechanisms like gaugino-squark condensation induce
superpotentials of the type {\it (a)} (see eq.(\ref{F18})) was
given\cite{Kim,Casas}. In ref.\cite{Narain}
the same kind of superpotential was obtained using pure gaugino condensation.
It was used the fact that in some classes of four-dimensional Superstrings
(orbifolds) a possible $H_1H_2$ dependence may appear in $f$ at one-loop.
The alternative mechanism {\it (b)} in which there is an
extra term in the K\"ahler
potential (see eq.(\ref{F9})) originating a $\mu$-term
is also naturally present in some large classes of four-dimensional
Superstrings\cite{KapluLouis,Dieter,Narain}.

In Superstring theory, neither the kinetic terms are in general canonical
nor the couplings ($Y_{ijk}$, $\lambda$, $Z$) and the mass term ($\mu$)
are independent of hidden fields. However, as we will see in subsect.3.2,
it is still possible to obtain (the phenomenologically desirable) universal
soft terms in the so-called dilaton-dominated limit\cite{KapluLouis,BIM}.
This limit is not only interesting because of that, but also because it is
quite model independent (i.e. for any compactification scheme the results
for the soft terms are the same). It is also remarkable that in this
limit once again the value of $B$ for the two mechanisms {\it (a)}, {\it (b)}
coincides with that of eq.(\ref{F323}). If, alternatively, we just
assume that a small ($\sim M_W$) dilaton-independent mass $\mu$ is
present in the superpotential, then the result for $B$ is now
given\cite{BIM} by eq.(\ref{F15}) as in the case of canonical kinetic terms.

We will study in detail this issue in the next section.


\section{Soft terms from Superstring theory}
\vspace{-0.7cm}
\subsection{General structure of soft terms}

\noindent Following the Introduction
and the notation of sect.2, the hidden and
observable sectors are given by $h_l = S, T_m$ and $\phi_i$ respectively.
In order to proceed we need to make some simplifying assumptions. We will
comment below what
changes are to be expected if the assumptions are relaxed.
Amongst the moduli $T_m$ we will concentrate on the overall modulus
$T$ whose classical value gives the size of the manifold. Apart from
simplicity, this modulus is the only one which is always necessarily present
in any $(0,2)$ (but left-right symmetric) 4-D Superstring.
 We believe that
studying the one modulus case is enough to get a feeling of the
most important physics of soft terms. Anyway, the case with several moduli
will be analyzed (for orbifolds) in subsect.3.3.
We will disregard for the moment any mixing between the $S$ and $T$ fields
kinetic terms. In fact this is strictly correct in all 4-D Superstrings at
tree level. However, it is known that this type of mixing may arise at one
loop level in some cases. On the other hand, these are loop effects
which should
be small and in fact can be easily incorporated in the analysis in some simple
cases (orbifolds) as shown in subsection 3.3.

Under the above conditions, the functions which appear in $K$ and $W$
eqs.(\ref{F9},\ref{F10}) have the following general expressions:
\begin{eqnarray}
K_0\ =\ -log(S+S^*)\ +\ K_0(T,T^*)\ \ ,\ \ K_i\ =\ K_i(T,T^*)\ \ ,\ \
Z\ =\ Z(T,T^*) \ ,
\label{F24} \\
W_0\ =\ W_0(S,T)\ \ ,\ \ \mu\ =\ \mu(S,T)\ \ ,\ \ Y_{ijk}\ =\ Y_{ijk}(T) \ .
\label{F25}
\end{eqnarray}
These confirm the above comment that in Superstring theory the kinetic terms
are non-canonical and the couplings are hidden field dependent.
The tree-level expression for $f_a$ for any four-dimensional
Superstring, as
mentioned in the Introduction, is well known,
$f_a=k_aS$, where $k_a$ is the Kac-Moody level of the gauge factor. Normally
(level one case) one takes $k_3=k_2=\frac{3}{5}k_1=1$. Since a possible
$T$ dependence may appear at one-loop\cite{IN}, then in general
\begin{eqnarray}
f_a(S,T)\ =\ k_aS + f_a(T)\ ,
\label{F26}
\end{eqnarray}
where we assume that other possible chiral fields do not contribute to
SUSY-breaking. It is important to stress that this gauge kinetic function
does not get further renormalized beyond one-loop and that it is therefore
an exact expression at all orders\cite{higher}.
Finally, the cosmological constant is (see eq.(\ref{F1}))
\begin{eqnarray}
V_0\ =\ {G_0}_S^S |F_0^S|^2\ +\ {G_0}_T^T |F_0^T|^2 \ -\ 3 e^{G_0}\ .
\label{F27}
\end{eqnarray}
Of course,
the first two terms in the right hand side of eq.(\ref{F27}) represent the
contributions of the $S$ and $T$ auxiliary fields,
$F_0^S=e^{G_0/2}({G_0}^S_S)^{-1} G_0^S$ and
$F_0^T=e^{G_0/2}({G_0}^T_T)^{-1} G_0^T$.

As we will show below, it is important to know what field, either $S$ or $T$,
plays the predominant role in the process of SUSY-breaking. This
will have relevant consequences in determining
the pattern of soft terms, and therefore the spectrum of physical
particles. That is why
it is very useful to  define an angle $\theta $
in the following way\cite{BIM}  (consistently with eq.(\ref{F27})):
\begin{eqnarray}
({G_0}^S_S)^{1/2}\ F_0^S\ &=&\ {\sqrt 3}C m_{3/2}\ e^{i\alpha _S}sin\theta\ ,
\nonumber\\
({G_0}^T_T)^{1/2}\ F_0^T\ &=&\ {\sqrt 3}C m_{3/2}\ e^{i\alpha _T}cos\theta\ ,
\label{F28}
\end{eqnarray}
where $\alpha_S,\alpha_T$ are the phases of $F_0^S$ and  $F_0^T$, and the
constant $C$ is defined as follows:
\begin{eqnarray}
C^2\ =\ 1\ +\ {{V_0}\over {3m_{3/2}^2}}\ .
\label{F29}
\end{eqnarray}
If the cosmological constant $V_0$ is assumed to vanish, one has $C=1$, but we
prefer for the moment to leave it undetermined. As we already mentioned
below eq.(\ref{F12}), the way one deals with the
cosmological constant problem is important.

Notice that, with the above assumptions, the goldstino field which is swallowed
by the gravitino in the process of supersymmetry breaking is proportional
to
\begin{eqnarray}
{\tilde {\eta }}\ =\ sin\theta \ {\tilde S}\ +\ cos\theta \ {\tilde T}\ ,
\label{F30}
\end{eqnarray}
where ${\tilde S}$ and ${\tilde T}$ are the canonically normalized
fermionic partners of the
scalar fields $S$ and $T$ (we have reabsorbed here the phases by redefinitions
of the fermions ${\tilde S},{\tilde T}$). Thus the angle defined above
may be appropriately termed {\it goldstino angle} and has a clear
physical interpretation as a mixing angle.

Then it is straightforward to compute the general form of the soft
terms\cite{BIM,BIMS}
\begin{eqnarray}
M_a &=&  {\sqrt 3}C m_{3/2}
\left[{{k_a ReS}\over {Ref_a}} e^{-i\alpha _S}sin\theta \
 + \ e^{-i\alpha _T}cos\theta \ {{f_a^T({G_0}_T^T)^{-1/2}}\over {2 Ref_a}}
\right]\ ,
\label{F31}\\
m_i^2 &=& 2m_{3/2}^2\ (C^2-1) \ +\ m_{3/2}^2C^2[1 +N_i(T,T^*)cos^2\theta ]\ ,
\nonumber\\
N_i(T,T^*) &=& {3 \over {{K_0}_T^T} }
\left( { { {K_i}_T {K_i}^T }
\over { {K_i}^2 }  }  \ -\
{ {{K_i}^T_T}\over  {K_i}  } \right)
\ = \
-3{{ (log K_i)^T_T} \over {{K_0}_T^T} }\ ,
\label{F32}\\
A_{ijk} &=& -{\sqrt 3}Cm_{3/2}[e^{-i\alpha _S}sin\theta \  +
\ e^{-i\alpha _T} cos\theta \; \omega_{ijk}(T,T^*)]\ ,
\nonumber\\
{\omega _{ijk}(T,T^*)} &=& ({K_0}_T^T)^{-1/2}
\left( {\sum _{p=i,j,k}} {{K_p^T}\over {K_p}}
\ -\ K_0^T\ -\ {{Y_{ijk}^T}\over {Y_{ijk}}} \ \right)\ ,
\label{F33}\\
B &=& \frac{1}{\widehat{\mu}}
      \left(  B_Z \; m_{3/2} X Z + B_{\mu} \; \mu \frac{W_0^*}{|W_0|} e^{K_0/2}
\right) (K_{H_1} K_{H_2})^{-1/2}\ ,
\nonumber\\
B_{\mu} &=& m_{3/2}\left[\ -1\ -C{\sqrt 3}e^{-i\alpha _S}sin\theta
\left(1-{{\mu ^S}\over {\mu }}(S+S^*)\right)\right.\nonumber\\
&+&\left. C{\sqrt 3}e^{-i\alpha _T}cos\theta
({K_0}_T^T)^{-1/2}\left( K_0^T +\ {{\mu ^T}\over {\mu }}
-{{{K}_{H_1}^T}\over {{K}_{H_1}}}
-{{{K}_{H_2}^T}\over {{K}_{H_2}}}\right)\right]\ ,
\nonumber\\
B_Z &=& \frac{m_{ 3/2}}{X} \left\{(3C^2-1) +
C {\sqrt 3}e^{-i\alpha _T}cos\theta
({K_0}_T^T)^{-1/2}\left({{Z^T}\over Z}-{{{K}_{H_1}^T}\over
{{K}_{H_1}}}-{{{K}_{H_2}^T}\over {{K}_{H_2}}}\right)\right.\nonumber\\
&-&C{\sqrt 3}e^{i\alpha _T}cos\theta ({K_0}_T^T)^{-1/2}{{Z_T}\over Z}
\nonumber\\
&+&\left. C^2 3({K_0}_T^T)^{-1}
cos^2\theta \left[{{Z_T}\over Z}\left({{{K}_{H_1}^T}\over
{{K}_{H_1}}}+{{{K}_{H_2}^T}\over {{K}_{H_2}}}\right)\
-\ {{Z^T_T}\over Z}\right]\right\}\ ,
\nonumber\\
X &\equiv& 1- C \sqrt{3} e^{i \alpha_T} \cos\theta
({K_0}^T_T)^{-1/2} \frac{Z_T}{Z}\ ,
\nonumber\\
\widehat{\mu} &=& \left[m_{3/2} X Z +
 \mu \frac{W_0^*}{|W_0|} e^{K_0/2} \right] (K_{H_1} K_{H_2})^{-1/2}\ ,
\label{F34}
\end{eqnarray}
%
where $Ref_a$ are the inverse squared gauge coupling constants
at the string scale.
Notice that we only include one-loop effects in the computation of the
soft gaugino masses. The motivation is the following:
one-loop corrections to the
$f_a$ function are known explicitely in some four-dimensional strings,
and moreover higher-loop corrections are vanishing as mentioned before, whereas
computing the one-loop-corrected bosonic soft terms would
require knowledge of the one-loop-corrected K\"ahler
potential, whose form is not available in the general case.
However, for some orbifolds models, well motivated conjectures give the
form of the one-loop-corrected K\"ahler potential: in particular,
mixing appears between the $S$ and $T$ kinetic terms. We
will take into account such corrections in subsect.3.3.
They are normally negligible, but may be
important for small $\sin\theta$, as we will see in specific cases.

The above expressions become much simpler in specific four-dimensional
Superstrings and/or in the large-T limit. This is the case for instance of
the formula for
$N_i(T,T^*)$ which looks complicated but it becomes very simple.
$N_i$ is related to the curvature of the K\"ahler manifold parametrized by the
above K\"ahler potential. For manifolds of constant curvature (like in
the orbifold case) the $N_i$ are constants, independent of $T$. More precisely,
they correspond to the modular weights of the charged fields, which are
normally negative integer numbers. In more complicated
four-dimensional Superstrings like those based on Calabi-Yau manifolds, the
$N_i(T,T^*)$ functions are complicated expressions in which world-sheet
instanton effects play an important role. In the case of $(2,2)$ Calabi-Yau
manifolds, for the large $T$ limit it turns out that
$N_i(T,T^*)\rightarrow -1$. We will come back to the evaluation of
the $N_i$ in specific Superstring models later on. Anyway,
the explicit dependence of the soft masses on the $N_i$
of each particle may produce in general
a lack of universality\cite{IL,Beatriz2}. This is
relevant for the issue of FCNC.
For an extended discussion on this point see section 5 of ref.\cite{BIM}
and refs.\cite{Pokorski,Nir}.

The soft terms obtained in the previous analysis
are in general complex.
Notice that if $S$ and $T$ fields acquire complex vacuum expectation
values, then the phases $\alpha _S,\alpha _T$ associated
with their auxiliary fields can be non-vanishing and
the functions $\omega _{ijk}(T,T^*)$, $f_a (T)$, etc.
can be complex. The analysis of this situation in connection with the
experimental limits on CP-violating effects like an electric dipole moment
for the neutron (EDMN)
can be found in section 4 of ref.\cite{BIM} and ref.\cite{Choi}.

Finally,
the above analysis has shown that the different soft SUSY-breaking terms
have all an explicit dependence on $V_0$, i.e. the cosmological constant,
which is contained in $C$.
We have to face this fact and do something
about it\footnote{It is worth noticing that
general properties which are independent
of the value of the cosmological constant can still be
found (see subsect.3.3 of ref.\cite{BIM}).}.
We cannot just simply ignore it, as it is often done, since
the way we deal with the cosmological constant problem has a bearing
on measurable quantities like scalar masses.
For an extended discussion on this point see sections 6 and 8 of ref.\cite{BIM}
and refs.\cite{KN,Park,Ferrara}.

As already explained at the end of subsect.2.2,
the superpotential eq.(\ref{F18}) which provides a possible solution to the
$\mu$ problem can naturally be obtained in the context of
Superstring theory with $\lambda=\lambda(T)$ in general\cite{Casas,Narain}.
So with this solution to the
$\mu$ problem, $B_{\mu}$ in eq.(\ref{F34}) gives\cite{BIMS}
\begin{eqnarray}
B_{\lambda}\ =m_{3/2} \left[ (3C^2-1) + C{\sqrt 3}e^{-i\alpha _T}cos\theta
({K_0}_T^T)^{-1/2}\left(\ {{\lambda ^T}\over {\lambda }}
-{{{K}_{H_1}^T}\over {{K}_{H_1}}}
-{{{K}_{H_2}^T}\over {{K}_{H_2}}}\right)\right]\ .
\label{F35}
\end{eqnarray}
The alternative mechanism {\it (b)}
in which there is an extra term in the K\"ahler
potential (see eq.(\ref{F9})) originating a $\mu$-term
is also naturally present in some large classes of four-dimensional
Superstrings. Indeed, in the case of some orbifold models and the large-T
limit of Calabi--Yau compactifications one
expect\cite{KapluLouis,Dieter,Narain}
\begin{eqnarray}
Z(T,T^*) \simeq \frac {1}{T+T^*}\ .
\label{F36}
\end{eqnarray}

\subsection{The $sin\theta =1$ (dilaton-dominated) limit}

\noindent Before going into specific classes of Superstring models,
it is worth studying the
interesting limit $\sin\theta =1$, corresponding to the case where
the dilaton sector is the source of all
the SUSY-breaking\cite{KapluLouis,BIM} (see
eq.(\ref{F28})). Since the
dilaton couples in a universal manner to all particles, {\it this limit is
quite model independent}.
Using eqs.(\ref{F31},\ref{F32},\ref{F33},\ref{F34},\ref{F35})
one finds the following simple expressions for the
soft terms:
\begin{eqnarray}
M_a\ &=&\ {\sqrt 3}Cm_{3/2}{{k_a ReS}\over {Ref_a}}e^{-i\alpha _S}\ ,
\nonumber\\
m_i^2\ &=&\ C^2 m_{3/2}^2\ +\ 2m_{3/2}^2(C^2-1)\ ,
\nonumber\\
A_{ijk}\ &=&\ -{\sqrt 3}Cm_{3/2}e^{-i\alpha _S}\ ,
\nonumber\\
B_{\mu }\ &=&\ m_{ 3/2}\left[ -1\ -\ {\sqrt 3}Ce^{-i\alpha _S}
\left(1-{{\mu ^S}\over {\mu }}(S+S^*)\right)\right]\ ,
\nonumber\\
B_Z\ &=&B_{\lambda}=\ m_{ 3/2}( 3C^2-1) \  ,
\label{F37}
\end{eqnarray}
where the scalar masses and the $A$-terms are universal,
whereas the gaugino masses may be slightly non-universal
since non-negligible threshold effects might be present.

Notice that the expressions for $m_i$ and $B_{\lambda}(B_{Z})$, using
eq.(\ref{F29}), coincide
with the ones obtained in SUGRA models with
canonical kinetic terms for the matter fields eqs.(\ref{F12}) and
(\ref{F21})(eq.(\ref{F23})). Notice also that the expression
for $B_{\mu }$ obtained in the limit
$\mu^S=0$ coincides with eq.(\ref{F15}).

It is obvious that this
limit $\sin \theta=1$ is quite predictive.
For a vanishing cosmological constant (i.e.
$C=1$), the soft terms are in the ratio $m_i:M_a:A=1:{\sqrt 3}:{-\sqrt 3}$
up to small threshold effect corrections (and neglecting phases). This will
result in definite patterns for the low-energy particle
spectra\cite{Barbieri,BIM} as we will see below.

\subsection{Computing soft terms in specific Superstring models}

\noindent In order to obtain more concrete
expressions for the soft terms one has to
compute the functions $N_i(T,T^*)$, $\omega _{ijk}(T,T^*)$ and $f_a(S,T)$. In
order
to evaluate these functions one needs a minimum of information about the
K\"ahler potential $K$, the structure of Yukawa couplings $Y_{ijk}(T)$ and the
one-loop threshold corrections $f_a(T)$. This type of information is only
known for some classes of four-dimensional Superstrings which deserve special
attention. We will thus concentrate here on two large classes of models:
the large-$T$ limit of Calabi-Yau compactifications\cite{Horowitz} and orbifold
compactifications\cite{Harvey}. We will describe the general pattern of soft
terms in
these large classes of Superstring models in turn.

\subsubsection{Orbifold compactifications}

\noindent In the case of orbifold four-dimensional Superstrings the K\"ahler
potential
has the general form (for small $|\phi_i|$)\cite{Witten,FKP,DKL1}
\begin{eqnarray}
K\ =\ -log(S+S^*)\ -3 log(T+T^*)\ +\sum _i (T+T^*)^{n_i} \phi_i \phi_i^* \ ,
\label{F38}
\end{eqnarray}
where the $n_i$ are normally negative
integers, sometimes called modular weights
of the matter fields (see ref.\cite{IL} for a classification of possible
modular weights of charged fields in orbifolds). For example, in the case
of $Z_N$ orbifolds the possible modular weights of matter fields are
$-1,-2,-3,-4,-5$.
Fields belonging to the untwisted sector have $n_i=-1$.
Fields in twisted sectors of the orbifold but without
oscillators have usually modular weight $n_i=-2$
(twisted associated to unrotated planes
of the underlying six-torus have $n_i=-1$)
and those with oscillators have $n_i\leq -3$.
It is important to remark that, unlike the case of smooth Calabi-Yau models,
the above $T$ dependence does not get corrections from world-sheet
instantons and
is equally valid for small and large $T$. In fact, general orbifold
models have a symmetry ("target-space duality") which relates\cite{FLT}
small to large
$ReT$. This is a discrete infinite subgroup of $SL(2,{\bf R})$ in which
$T$ plays the role of modulus. For the overall field $T$ here considered,
the target-space duality group will be either
the modular group $SL(2,{\bf Z})$ or a subgroup of it
(in some cases, if quantized Wilson lines are present). Under $SL(2,{\bf Z})$
the modulus transforms like
\begin{eqnarray}
T\ \rightarrow \ {{aT-ib}\over {icT+d}} \ \ ;\ \ ad-bc\ =\ 1\ ,\ a,b,c,d\in
{\bf Z}\ ,
\label{F889}
\end{eqnarray}
the dilaton $S$ field is invariant at tree-level and the matter fields
transform like
\begin{eqnarray}
C_i\ \rightarrow \ \ (icT+d)^{n_i}\ C_i\
\label{F888}
\end{eqnarray}
up to constant matrices which are not relevant for the present analysis.
Eq.(\ref{F888})
explains why the integers $n_i$ are called modular weights. With the above
transformation properties the $G$ function is modular invariant (if the
superpotential $W$ has modular weight $-3$).

The threshold correction functions $f_a(T)$ (due to the contribution of
massive string states in the loops) have been computed for
$Z_N$ and $Z_N\times Z_M$ orbifolds only for the $(2,2)$ case
in refs.\cite{Kaplunovsky,DKL2,Louis,DFKZ,Cual}
although they are expected to be valid for more general cases (see e.g.
ref.\cite{IL} for a general discussion on this point).
The result has the form\footnote{For the case when the underlying six
dimensional torus lattice is not assumed to decompose into a direct
sum of a four-dimensional and a two-dimensional sublattice, with the
latter lying in a plane left fixed by a set of orbifold twists, see
\cite{Stieberger,Bailin}.}
\begin{eqnarray}
f_a(T)\ =\ -\ {1\over {16\pi ^2}}(b_a'-k_a\delta _{GS})\ log\ \eta ^4(T)\ \ ,
\label{F39}
\end{eqnarray}
where $\eta (T)$ is the well known Dedekind function which admits the
representation
\begin{eqnarray}
\eta (T)\ =\ e^{-\pi T/12}\ \Pi _{n=1}^{\infty}(1-e^{2\pi nT})\ .
\label{F40}
\end{eqnarray}
In eq.(\ref{F39})
$k_a$ is the Kac-Moody level of the gauge group $G_a$, $\delta
_{GS}
$ is
a group independent (but model-dependent) constant and
\begin{eqnarray}
b_a'\ =\ -3C(G_a)\ +\ \sum _i T_a(\phi_i)(3 + 2n_i)
=\ b_a\ +\ 2\sum _i T_a(\phi_i)(1 + n_i)\ \ ,
\label{F41}
\end{eqnarray}
where
$C(G_a)$ denotes the quadratic Casimir in the adjoint representation
of $G_a$,  $T_a(\phi_i)$ is defined by $Tr(T^{\alpha} T^{\beta} )
=T_a(\phi_i){\delta}^{\alpha \beta}$ ($T^{\alpha}=$ generators of $G_a$
in the $\phi_i$ representation)
and the sum runs over all the massless charged chiral fields.
Notice that $b_a$ is nothing but the $N=1$ one-loop $\beta $-function
coefficient
of the $G_a$ gauge coupling and that $b_a'=b_a$ when all matter fields have
modular weights $n_i=-1$ (as for untwisted states). While $b_a'$ may be
computed
in terms of the effective low-energy degrees of freedom of the theory,
$\delta _{GS}$ is a model-dependent quantity (usually a negative integer
in the case of the overall modulus $T$).
Its presence is associated to the cancellation of
the one-loop duality
anomalies of the theory.

Using
eqs.(\ref{F26},\ref{F31},\ref{F32},\ref{F33},\ref{F34},\ref{F35},\ref{F36})
and the above expressions for $K$ and $f_a(T)$
one obtains the soft terms\cite{BIM,BIMS}
\begin{eqnarray}
M_a &=& {\sqrt 3}C m_{3/2}
\left({{k_a ReS}\over {Ref_a}} e^{-i\alpha _S}sin\theta\right.\nonumber\\
&+& \left. e^{-i\alpha _T}cos\theta {{(b_a'-k_a\delta _{GS})(T+T^*)
{\hat {G_2}}(T,T^*)}\over{32\pi^3{\sqrt 3}Ref_a}}\right)\ ,
\label{F42}\\
m_i^2 &=& m_{3/2}^2C^2(1\ +\ n_icos^2\theta )\ +\ 2m_{3/2}^2(C^2-1)\ ,
\label{F43}\\
A_{ijk} &=& -{\sqrt 3} C m_{3/2}[ e^{-i\alpha _S}sin\theta\ +\
 e^{-i\alpha _T} cos\theta\ \omega _{ijk}(T,T^*)\ ]\ ,
\nonumber\\
\omega _{ijk}(T,T^*) &=& {1\over {\sqrt 3}}\left(3+n_i+n_j+n_k\ -(T+T^*)
{{Y_{ijk}^T}\over {Y_{ijk}}}\ \right)\ ,
\label{F44}\\
B_{\mu } &=& m_{ 3/2}\left[\ -1\ -\ C{\sqrt 3}e^{-i\alpha _S}sin\theta
\left(1-{{\mu ^S}\over {\mu }}(S+S^*)\right)\right.\nonumber\\
&-& \left.C e^{-i\alpha _T}cos\theta  \left(3+n_{H_1}+n_{H_2}
-\ {{\mu ^T}\over {\mu }}(T+T^*)\right)\right] \ ,
\label{F444}\\
B_{\lambda} &=& m_{3/2}\left[\ (3C^2-1) - Ce^{-i\alpha _T}cos\theta
\left(n_{H_1}+n_{H_2}-{{\lambda ^T}\over {\lambda }}(T+T^*)\right)\right],
\label{F445}\\
B_Z\ &=& \frac{m_{ 3/2}}{X} [(3C^2-1) -
C cos\theta
(e^{-i\alpha_T}(n_{H_1}+n_{H_2}+1)-e^{i\alpha _T})
\nonumber\\
&-& C^2 cos^2\theta (2+n_{H_1}+n_{H_2})]\ ,
\nonumber\\
X &\equiv& 1+ C e^{i \alpha_T} cos\theta \ ,
\label{F45}
\end{eqnarray}
where ${\hat {G_2}}$ is the non-holomorphic Eisenstein function which may be
defined by ${\hat {G_2}}=G_2(T)-2\pi/(T+T^*)$. Here $G_2$ is the holomorphic
Eisenstein form which is related to the Dedekind function by
$G_2(T)=-4\pi(\partial \eta (T)/\partial T)(\eta (T))^{-1}$.
In fact, using eq.(\ref{F31})
one gets eq.(\ref{F42}) with $G_2$
instead of ${\hat {G_2}}$. However, one gets the
complete modular invariant result in
eq.(\ref{F42}) when one includes the one-loop contribution of
massless
fields (see ref.\cite{IL}) and the one-loop (Superstring) correction to the
K\"ahler
potential (see ref.\cite{BIMS}).

Notice the explicit dependence of the soft masses on the modular
weights of each particle. As mentioned in subsection 3.1, this lack of
universality may be relevant for the issue of FCNC.

The $A$-parameters
depend on the modular weights
of the particles appearing in the Yukawa coupling.
Moreover, the last term in $\omega_{ijk}$, eq.(\ref{F44}), drops
for Yukawa couplings
involving either untwisted fields or (for large $T$) twisted fields associated
to the $same$
fixed point. The reason is that the Yukawa couplings
are constants or tend exponentially to constants, respectively\cite{Hamidi}.
This simplify
the phenomenological analysis
since the relevant couplings will be of
this type.
(The $A$ term which is
relevant to electroweak symmetry breaking is the one associated
to the top-quark Yukawa coupling. If the fields are twisted
they should be associated to the same fixed point in order
to obtain the largest possible value of the coupling, otherwise
it would be exponentially suppressed\cite{Faustino}.)

\vspace{0.2cm}

\noindent {\it One-loop (Superstring) corrections}

\vspace{0.2cm}

\noindent As discussed in
subsection 3.1 there is a slight inconsistency in using one-loop formulae
for the gaugino masses whereas for the other
soft terms $m_i$, $A$ and $B$ we use only the tree level result. In fact this
is not that
important since normally the one-loop corrections are small for those terms.
Anyway, they may be evaluated knowing the one-loop (Superstring) corrections to
the
K\"ahler potential. General arguments applicable to orbifolds allow us to write
the one-loop corrected K\"ahler potential for orbifolds by making the
replacement\cite{DFKZ}
\begin{eqnarray}
S+S^*\ \longrightarrow \ \ Y=S+S^*\ -\
{{\delta _{GS}}\over {8\pi ^2}}log(T+T^*)
\label{F46}
\end{eqnarray}
in eq.(\ref{F38}). $\delta _{GS}$ {\it measures the
amount of one-loop mixing between the $S$ and $T$ fields} in the K\"ahler
potential. This one-loop mixing term with coefficient $\delta _{GS}$
generalizes
the Green--Schwarz mechanism\cite{Schwarz} and cancels anomalies of the
underlying non-linear $\sigma$-model\cite{DFKZ,Louis,Ovrut}, which are
described by triangle diagrams with two external gauge bosons and several
external moduli fields $T$. This type of mixing is expected to be present in
generic
four-dimensional strings and not only in the orbifold case. Now, for the
one-loop K\"ahler potential
to transform in the required way under target-space duality transformations
the dilaton has to acquire a non-trivial modular transformation behaviour
at the one-loop level\cite{DFKZ}
\begin{eqnarray}
S\ \rightarrow \ \ S-{{\delta _{GS}}\over {8\pi ^2}}log(icT+d) \ .
\label{F988}
\end{eqnarray}
The expression (\ref{F27}) for the VEV of the scalar
potential gets modified as follows\cite{BIM}:
\begin{eqnarray}
V_0 &=& e^{G_0}  Y^2 |G_0^S|^2 +  e^{G_0}
{ {(T+T^*)^2} \over  3} \left(1- { {\delta _{GS}} \over {24\pi^2 Y} }
\right)^{-1}
 \left|G_0^T+  { {\delta _{GS}} \over {8\pi^2(T+T^*)} } G_0^S\right|^2 - 3
e^{G_0} \nonumber\\
 & =&\ {1 \over {Y^2}}
 \left|F_0^S - { {\delta _{GS}} \over {8\pi^2(T+T^*)} } F_0^T\right|^2 +
{ 3 \over  {(T+T^*)^2} } \left(1- { {\delta _{GS}} \over {24\pi^2 Y} } \right)
|F_0^T|^2
- 3 e^{G_0}\ .
\label{F47}
\end{eqnarray}
Analogously to eq.(\ref{F27}), we have written $V_0$ also in terms of the $S$
and $T$
auxiliary fields $F_0^S$ and $F_0^T$, whose 'mixed' relation with  $G_0^S$
and $G_0^T$
can be easily read off from eq.(\ref{F47}) itself (the equality holding
term by term).

In the present case, we modify the definition (\ref{F28}) of the $\theta$ angle
(and the phases) in the following way (consistently with eq.(\ref{F47})):
\begin{eqnarray}
 {1 \over Y}\left( F_0^S - { {\delta _{GS}} \over {8\pi^2(T+T^*)} }
F_0^T\right)
&=&\ {\sqrt 3}C\ m_{3/2} e^{i\alpha _S}sin\theta \ ,
\nonumber\\
{{\sqrt 3} \over  {T+T^*} } \left(1- { {\delta _{GS}} \over
{24\pi^2 Y} } \right)^{1/2} F_0^T
&=&\ {\sqrt 3} C\ m_{3/2} e^{i\alpha _T}cos\theta \ .
\label{F48}
\end{eqnarray}
Notice that eqs.(\ref{F47}) and (\ref{F48}) reduce to eqs.(\ref{F27})
and (\ref{F28}) in the limit $\delta _{GS}=0$, as they should.

After computing again the soft terms, one finds that the
resulting expressions can in practice
be obtained\cite{BIM} from the previous formulae
(\ref{F42},\ref{F43},\ref{F44},\ref{F445},\ref{F45})
by making the replacements (\ref{F46}) and
\begin{eqnarray}
cos\theta \ \longrightarrow \ \left(1-{ {\delta _{GS}}\over {24\pi^2 Y}
}\right)^{-1/2}
\ cos\theta
\label{F49}
\end{eqnarray}
(without changing $\sin\theta$).
The same happens with the formulae for the $B_{\mu}$ parameter
eq.(\ref{F444}) but also including\cite{BIMS} a term
$-\frac {\delta _{GS}}{8\pi^2} \frac {\mu^S}{\mu}$
inside the parenthesis that multiplies to $cos\theta$.

We anticipate that the corrections due to $S$--$T$ mixing are normally
negligible for not too large  $\delta _{GS}$. However {\it they turn out
to be important in the $\sin\theta \rightarrow 0$ limit} in which all
tree level masses become small and one-loop effects cannot be neglected.

Notice that now  $F_S$ is non-vanishing when $\sin\theta \rightarrow  0$,
differently from the case without $S-T$ mixing (see eq.(\ref{F28})).
However, if $\delta _{GS}$ is not too large, this limit still corresponds
to a modulus dominated SUSY-breaking.
Indeed, from eq.(\ref{F48}) one obtains
\begin{eqnarray}
|{{F_0^S} \over{F_0^T}}| = {|{\delta_{GS}}| \over{8\pi^2 (T+T^*)} } << 1  \ .
\label{F50}
\end{eqnarray}
%


\vspace{0.2cm}
\noindent {\it The case with several moduli}
\vspace{0.2cm}

\noindent In the case with several
moduli ($T_m$) the situation is more cumbersome
and one is forced to define new goldstino angles. This was first done in
section 8 of ref.\cite{BIM} in a different context (extra matter fields).
Following this line, the VEV of the scalar potential (see eq(\ref{F27}))
gets modified as
\begin{eqnarray}
V_0\ =\ {G_0}_S^S |F_0^S|^2\ +\ \sum _m {G_0}_{T_m}^{T_m} |F_0^{T_m}|^2 \ -\ 3
e^{G_0}
\label{F277}
\end{eqnarray}
and eq.(\ref{F28}) is modified to\cite{Unpub,Japoneses}
\begin{eqnarray}
({G_0}^S_S)^{1/2}\ F_0^S\ &=&\ {\sqrt 3}C m_{3/2}\ e^{i\alpha _S}
sin\theta \ ,
\nonumber\\
({G_0}^{T_m}_{T_m})^{1/2}\ F_0^{T_m}\ &=&\ {\sqrt 3}C m_{3/2}\
e^{i\alpha _{T_m}}cos\theta \Delta_m \ .
\label{F95}
\end{eqnarray}
where $\sum_m  \Delta_m^2\ =1$.
For instead, for $m=1,...,4$ (this is e.g. the case of some $Z_N$ and
$Z_N \times Z_M$
orbifolds with three
diagonal (1,1) moduli ($T_1,T_2,T_3$) and one (2,1) moduli ($T_4=U$)),
three new goldstino angles are
necessary: $\Delta_1=cos\theta_1, \Delta_2=sin\theta_1 cos\theta_2,
\Delta_3=sin\theta_1 sin\theta_2 cos\theta_3, \Delta_4=sin\theta_1
sin\theta_2 sin\theta_3$.
Now, with these definitions, and taking into account that eq.(\ref{F38})
is modified to
\begin{eqnarray}
K\ =\ -log(S+S^*)\ - \sum _m log(T_m+T_m^*)\ +\sum _i \phi_i \phi_i^* \prod _m
(T_m+T_m^*)^{n_i^m} \ ,
\label{F388}
\end{eqnarray}
the computation of the soft terms gives
\begin{eqnarray}
M_a &=& {\sqrt 3}Cm_{3/2}
\left({{k_a ReS}\over {Ref_a}}e^{-i\alpha _S}sin\theta\right.\nonumber\\
&+& \left.  cos\theta \sum_m e^{-i\alpha_{T_m}}
{{(b_a'^m-k_a\delta _{GS}^m)(T_m+T^*_m)
{\hat {G_2}}(T_m,T^*_m)}\over{32\pi ^3 Ref_a}} \Delta_m\right) \ ,
\nonumber\\
m_i^2 &=& m_{3/2}^2C^2\left(1\ +3 cos^2\theta\ \sum_m  n_i^m \Delta_m^2\right)\
+\ 2m_{3/2}^2(C^2-1) \ ,
\nonumber\\
A_{ijk} &=& -{\sqrt 3} C m_{3/2}\left[ e^{-i\alpha _S}sin\theta\ +\
cos\theta \sum_m e^{-i\alpha _{T_m}} \Delta_m \omega _{ijk}(T_m)\ \right] \ ,
\nonumber\\
\omega _{ijk}(T_m) &=& 1+n_i^m+n_j^m+n_k^m\ -(T_m+T^*_m)
{{Y_{ijk}^{T_m}}\over {Y_{ijk}}} \ ,
\label{F97}
\end{eqnarray}
where $n_i^m$ are the modular weights of the matter fields
and $b_a'^m=-C(G_a)+\sum_i T_a(\phi_i)(1+2n_i^m)$.
The overall modulus case ($T_1=T_2=T_3=T,T_4=0$) soft terms
eqs.(\ref{F42},\ref{F43},\ref{F44}) are recovered for
$\Delta_m=1/\sqrt3, \delta_{GS}^m=\delta_{GS}/3, Y_{ijk}^{T_m}=Y_{ijk}^{T}/3$,
m=1,2,3 and $\Delta_4=0$.
A more complete
analysis including the B-term, one-loop (Superstring) corrections,
phenomenological consequences, and a comparison with the overall modulus
($T$) case can be found in ref.\cite{BIMS}.

\subsubsection{Large-$T$ limit of Calabi--Yau compactifications}

\noindent Little is
known about the general form of the K\"ahler potential and couplings
of generic Calabi--Yau $(2,2)$ compactifications. Only a few examples
(most notably, the quintic in $CP^4$) have been worked out\cite{Candelas}
in some detail\cite{Dixonunpub}
and show formidable complexity due to the world-sheet instanton contributions
to the K\"ahler potential.
On the other hand, a few generic facts concerning these models are known
for the large-$T$ limit\cite{Dixonunpub,Lustcvetic}.
Large $T$, in practice, does not really mean
$T\rightarrow \infty$, since the world-sheet instanton corrections are
exponentially suppressed. For values $|T|\geq 2-3$ these world-sheet instanton
contributions can often be neglected and, in this sense these $T$-values are
already large. It is true that $|T|$ cannot be infinitely large, since
otherwise the quantum corrections to the gauge coupling constants (string
threshold corrections) may be too large and spoil perturbation theory. The
maximum allowed $|T|$ not spoiling perturbation theory is something which
is model dependent but is expected to be much bigger than one since, after
all, the threshold corrections are loop effects. In explicit orbifold
examples it was found in refs.\cite{ILR,IL} that $|T|\leq 20-30$ is enough
to remain in
the perturbative regime. When we talk about the large $T$ limit in what
follows we will thus assume $T$-values which do not spoil perturbativeness.
In this limit the K\"ahler potential $K$ gets a
particularly simple form:
\begin{eqnarray}
K(T\rightarrow \infty)\ =\ -log(S+S^*)\ -3 log(T+T^*)\ +\sum _i (T+T^*)^{-1}
\phi_i \phi_i^* \ .
\label{F1001}
\end{eqnarray}
Notice that, the resulting K\"ahler potential is analogous to the one obtained
in orbifold models eq.(\ref{F38}) with matter fields in the untwisted sector.
Therefore we can use
eqs.(\ref{F42},\ref{F43},\ref{F44},\ref{F444},\ref{F445},\ref{F45})
with $n_i=-1$ in order to obtain
$T\rightarrow \infty$ soft terms:
\begin{eqnarray}
M_a &=& {\sqrt 3}C m_{3/2}
{{k_a ReS}\over {Ref_a}} e^{-i\alpha _S}sin\theta\ ,
\label{F1002}\\
m_i^2 &=& m_{3/2}^2C^2 sin^2\theta\ +\ 2m_{3/2}^2(C^2-1)\ ,
\label{F1003}\\
A_{ijk} &=& -{\sqrt 3} C m_{3/2}[ e^{-i\alpha _S}sin\theta\ +\
 e^{-i\alpha _T} cos\theta\ \omega _{ijk}(T,T^*)\ ]\ ,
\nonumber\\
\omega _{ijk}(T,T^*) &=& -{{(T+T^*)}\over {\sqrt 3}}
{{Y_{ijk}^T}\over {Y_{ijk}}} \ ,
\label{F1004}\\
B_{\mu } &=& m_{ 3/2}\left[\ -1\ -\ C{\sqrt 3}e^{-i\alpha _S}sin\theta
\left(1-{{\mu ^S}\over {\mu }}(S+S^*)\right)\right.\nonumber\\
&-& \left.C e^{-i\alpha _T}cos\theta  \left(1-\ {{\mu ^T}\over {\mu
}}(T+T^*)\right)\right] \ ,
\label{F1005}\\
B_{\lambda} &=& m_{3/2}\left[\ (3C^2-1) + Ce^{-i\alpha _T}cos\theta
\left(2+{{\lambda ^T}\over {\lambda }}(T+T^*)\right)\right] \ ,
\label{F1006}\\
B_Z\ &=& \frac{m_{ 3/2}}{X} [(3C^2-1) + 2C cos\alpha_T cos\theta ] \ ,
\nonumber\\
X &\equiv& 1+ C e^{i \alpha_T} cos\theta \ ,
\label{F1007}
\end{eqnarray}
where we have ignored the possible one-loop corrections to these
formulae.

It can be further argued that, in the large $T$-limit,
the non-vanishing Yukawa
couplings tend (exponentially) to constants, as computed in specific examples.
Then one can
take $\omega _{ijk}\rightarrow 0$ in the mentioned limit. Of course, this
simplify the phenomelogocial analysis.

It is interesting to remark that in this large-$T$ limit of Calabi--Yau-type
compactifications
the results obtained for the soft scalar and gaugino masses
and A-parameters are quite similar to those in
eq.(\ref{F37}) obtained in a model-independent manner for $\sin\theta =1$. The
role of dimensionful parameter is played now by $m_{3/2}\sin\theta $ (for
$C=1$)
instead of simply $m_{3/2}$.
Thus {\it dilaton-dominated SUSY-breaking is not the only situation in which
universal soft scalar masses are obtained}, as the present model exemplifies.
Anyhow we point out that in the case with several moduli the situation might be
much more cumbersome and one is forced to define new goldstino angles (as we
did in the orbifold case in the previous subsection).
If we allow generic values for these angles we obtain a deviation from
the previous universal behaviour (see e.g. the soft scalar masses in
eq.(\ref{F97}) for the orbifold case).

It is also interesting
to notice how, for $C=1$,  all these terms
tend to zero at the same speed as $\sin\theta \rightarrow 0$, even for a finite
value of $m_{3/2}$. Indeed, for a very small $\sin\theta$ the gravitino mass
$m_{3/2}$ decouples from the SUSY-breaking soft terms and may become much
larger
than them.
However, for $\sin \theta =0$ the one-loop corrections to
the K\"ahler potential cannot probably be neglected (unfortunately,
the one-loop corrections to the K\"ahler potential
in the large-$T$ limit of Calabi--Yau compactifications are unknown),
and care should be taken before
getting any definite conclusion (see the case of one-loop orbifold corrections
discussed above).

The above statements concerning the large $T$-limit of Calabi-Yau
compactifications
are known to be true for $(2,2)$ models, which yield a gauge group $E_6\times
E_8$.
In order to make contact with the standard model one has to break this
structure with Wilson line gauge symmetry breaking and/or use $(0,2)$ type
compactifications. However, it is reasonable to expect that the general
structure in eq.(\ref{F1001})
will still apply in these more complicated cases and, hence,
eqs.(\ref{F1002},\ref{F1003},\ref{F1004},\ref{F1005},\ref{F1006},\ref{F1007})
will still hold.

Notice that the $\sin\theta\rightarrow 0$ limit of the large-$T$ Calabi--Yau
Superstrings is different from the "no-scale" supergravity models discussed in
the literature\cite{EllisCremmer}.
Although in both models (for $C$=1) one has at the
tree level $m_i=A=0$, the behaviour of the gaugino mass is totally different.
In the no-scale models the gaugino mass is non-vanishing and constitutes
the only source of SUSY-breaking whereas in the present class of
models the gaugino mass also vanishes.

\subsubsection{Supersymmetric-spectra expected in Superstring models}

\noindent The
formulae for soft terms written in the previous subsections may lead
to different phenomenological situations\cite{BIM} depending on, e.g.,
the phases,
the value of the cosmological constant, the possible values of the modular
weights of the particles, the ansatz for the $B$-parameter, etc. In what
follows we will assume vanishing phases and cosmological
constant\footnote{The impact of a non-vanishing tree-level
cosmological constant $V_0$ on SUSY-spectra has been
studied in ref.\cite{Park}.}.
The former is consistent with the experimental
limits on the EDMN and the latter with the experimental constraints in present
cosmology. In order to have $V_0=0$ one may assume that there
is {\it some yet undiscovered dynamics which
guarantees that $V_0=0$} at the minimum. In our context this implies
that the dynamics of the $S$ and $T$ superfields is such that their
auxiliary fields break supersymmetry with vanishing cosmological constant.
This could come about e.g. if some non-perturbative dynamics generates
an appropriate superpotential $W(S,T)$ with this $V_0=0$ property. Indeed,
such type of superpotentials can be constructed
(see ref.\cite{Cvetic,Ferrara}),
although the
physical origin of them is certainly obscure. Anyway, if one adopts this
philosophy one must set $V_0=0$ (or, equivalently, $C=1$) in all the
expressions for the soft terms.

First, let us compare the soft gaugino and scalar masses in the overall modulus
case. From eqs.(\ref{F42},\ref{F43}) these are given roughly by
\begin{eqnarray}
M_a^2 &=& 3 \ m_{3/2}^2 \ sin^2\theta \ ,
\label{F01}\\
m_i^2 &=& m_{3/2}^2(1\ +\ n_icos^2\theta )\ ;\ n_i=-1,-2,-3,... \ ,
\label{F02}
\end{eqnarray}
where we have neglected small threshold effect corrections ($ReS\simeq Ref_a$).
With respect to the one-loop correction (\ref{F46}), which
only appears in gaugino masses, it turns out to be also irrelevant
due to the fact that $\delta _{GS}$ is usually small.
In all orbifold models considered up to now $\delta _{GS}$
is of the same order of magnitude as the $b'$-coefficients appearing in the
model.
Both the one-loop ($S$-$T$ mixing) corrections eq.(\ref{F49})
and the one-loop
gaugino term proportional to $\cos\theta$ in eq.(\ref{F42}) are also neglected.
As was already mentioned
in subsection 3.3.1, the latter turn out to be numerically irrelevant for
not too small $\sin\theta$. This is usually the situation since
\begin{eqnarray}
cos^2\theta &\leq& {{1}\over {|n_i|}}
\label{F03}
\end{eqnarray}
in order to avoid negative mass-squared of the scalars of modular weight $n_i$
(see eq.(\ref{F02})). E.g. $n_i=-3$ implies $\sin\theta \geq 0.82$.
Thus the allowed values of $\sin\theta$ are relatively close to one and
{\it the dilaton is necessarily the dominant source of SUSY-breaking}.
Notice, however, that the moduli contribution
to SUSY-breaking is $not$ in general negligible.
Of course, for $|\sin\theta|=1$
one recovers the dilaton-dominated SUSY-breaking results.
Now, eqs.(\ref{F01},\ref{F02}) clearly imply that soft gaugino masses
(using the relation $\sin^2\theta=1-\cos^2\theta$)
are bigger than soft scalar masses
\begin{eqnarray}
M_a > m_i \ .
\label{F04}
\end{eqnarray}

Let us now discuss the predictions for the low-energy ($\sim M_Z$)
sparticle spectra. There are several particles whose mass
is rather independent of the details of $SU(2)_L\times U(1)_Y$ breaking and is
mostly given by the boundary conditions and the renormalization group running.
In particular, in the approximation that we will use
(neglecting all Yukawa couplings except the one of the top), that is
the case of the gluino $g$, all the squarks (except stops and left sbottom)
$Q_L=(u_L,d_L), u_L^c, d_L^c$ and all the sleptons $L_L=(v_L,e_L), e_L^c$.
For all these particles
one can write explicit expressions for the masses in terms of the gravitino
mass and $\sin\theta$.
\begin{eqnarray}
M_g^2(M_Z) \ &=& 9.8 \ M^2 = 29.4 \ m_{3/2}^2 \ sin^2\theta  \ ,
\label{F00004}\\
m_{Q_L}^2(M_Z) \ &=& m^2_{Q_L} + 8.3 \ M^2 = m_{3/2}^2 \ (1+n_{Q_L} cos^2\theta
+ 25 sin^2\theta) \ ,
\label{F00005}\\
m_{u_L^c,d_L^c}^2(M_Z) \ &=& m^2_{u_L^c,d_L^c} + 8 \ M^2 = m_{3/2}^2 \
(1+n_{u_L^c,d_L^c} cos^2\theta + 24 sin^2\theta) \ ,
\label{F00006}\\
m_{L_L}^2(M_Z) \ &=& m^2_{L_L} + 0.7 \ M^2 = m_{3/2}^2 \ (1+n_{L_L} cos^2\theta
+ 2 sin^2\theta) \ ,
\label{F00007}\\
m_{e^c_L}^2(M_Z) \ &=& m^2_{e^c_L} + 0.23 \ M^2 = m_{3/2}^2 \ (1
+n_{e^c_L} cos^2\theta
+ 0.7 sin^2\theta) \ ,
\label{F00008}
\end{eqnarray}
where the last term in
eqs.(\ref{F00005},\ref{F00006},\ref{F00007},\ref{F00008})
gives the effect of
gaugino loop contributions in the low-energy running. In the previous
formulae we have neglected the scalar potential D-term contributions which
are normally small and the contribution to the scalar mass RGEs of the
$U(1)_Y$ D-term. These may be found in eq.(9) of ref.\cite{Amanda}.
Now, the low-energy mass relations turn out to be
\begin{eqnarray}
m_l < m_q \simeq M_g .
\label{F09}
\end{eqnarray}
since the low-energy scalar masses are mainly determined
by the gaugino contributions. (This also implies that even with
non-vanishing $V_0$, these results are still maintained.) The slepton masses
are smaller than squark
masses because they do not feel the important gluino contribution.

Let us consider two examples in order to analyze the previous relations
in more detail.
In the first example we will assume
different (flavour-independent) modular weights
 for the different squark and sleptons within each generation. The
modular weights are chosen so that one can have appropriate large
string threshold corrections to fit the joining of gauge coupling
constants at a scale $\simeq 10^{16}$ GeV (see refs.\cite{ILR,IL}).
\begin{eqnarray}
n_{Q_L}=n_{d_L^c}=-1\ \ ,\ \ n_{u_L^c}=-2\ \ ,\ \
 n_{L_L}=n_{e_L^c}=-3 \ \ ,
n_{H_1}+n_{H_2}=-5,-4 \ .
\label{F0010}
\end{eqnarray}
The above values together with a $ReT\simeq 16$ lead
to good agreement for $\sin^2\theta _W$ and $\alpha _3$.
This scenario is also interesting
because it provides us with an explicit model with
non-universal scalar masses and shows the general features of
models with some of the modular weights different from $-1$.
For the sake of definiteness, in the following we will focus on the case
$n_{H_1}=-2, n_{H_2}=-3$. Other possible choices do not lead to significative
modifications in the phenomenological results.
For the masses of scalar particles with modular weights $-1,-2$ and $-3$,
eq.(\ref{F02}) gives respectively
\begin{eqnarray}
m_{-1}^2= m_{3/2}^2sin^2\theta \ ,\
m_{-2}^2= m_{3/2}^2(1-2cos^2\theta ) \ ,\
m_{-3}^2= m_{3/2}^2(1-3cos^2\theta ) \ .
\label{F0011}
\end{eqnarray}
Several comments are in
order. First of all the goldstino angle is constrained to
$\sin^2\theta \geq 2/3 $, otherwise
the scalars with modular weight $-3$ (the sleptons and the Higgs field
$H_2$ in the present case)
would get negative squared masses. For $| sin\theta | =1$ one
recovers the dilaton
dominated SUSY-breaking results. However, as one gets away from that value
important deviations
occur. The scalars with large (negative) modular weights get soft masses
substantially smaller
than those with e.g. $n_i=-1$. In particular, the sleptons are much lighter
than the
squarks already at the string scale for $|\sin\theta| \simeq 0.8$.
Therefore we have to give
the masses of the gluino, the squarks and the sleptons for a fixed
value of $\theta$.
In particular, we have chosen the smallest possible value
$\sin^2\theta = 2/3$.
For such value of $\theta$
the contribution of the moduli field $T$ to SUSY-breaking is substantial.
The mass ratios now turn out to be
\begin{eqnarray}
M_g:m_{Q_L}:m_{u_L^c}:m_{d_L^c}:m_{L_L}:m_{e_L^c}\ \simeq \
1:0.95:0.92:0.92:0.25:0.11 \ \ ,
\label{F0012}
\end{eqnarray}
where we have included the small contribution to the scalar mass RGEs of the
$U(1)_Y$ D-term.
The results are qualitatively similar to
those of the second scenario (see below), in spite of the different set of soft
scalar masses, because the low-energy scalar masses are mainly determined
by the gaugino contributions. The only exception is the $e_L^c$ mass,
which only feels the small $B$-ino contribution.
The similarity with eq.(\ref{F0009}) is clear (the results would be
still more similar for the other possible values of $\theta$).
Again, the masses do not contain
the scalar potential D-term contributions, which depend on the process of
symmetry breaking
and anyway are very small. In particular, concerning the possible shifts
of the slepton masses, similar remarks as in the second example apply (see
below).

In the second scenario we will assume that
all particles have modular weight $-1$, as
in the large-T Calabi--Yau limit and possible orbifold Superstring
scenarios. $M_a$ and $m_i$ scale, to a first approximation like
$m_{3/2}\sin\theta$ (see eqs.(\ref{F01},\ref{F02})) thus the
gluino and the squark and slepton masses are in the universal ratio
(at the $M_Z$ scale)
\begin{eqnarray}
M_g:m_{Q_L}:m_{u_L^c}:m_{d_L^c}:m_{L_L}:m_{e_L^c}\ \simeq\
1:0.94:0.92:0.92:0.32:0.24 \ \ .
\label{F0009}
\end{eqnarray}
Although squarks and sleptons have the same soft mass, at low-energy
the former are much heavier than the latter because of the gluino contribution
to the renormalization of their masses
(see eqs.(\ref{F00005},\ref{F00006},\ref{F00007},\ref{F00008})).
Actually, the result corresponds
to scalar masses without D-term contributions. The latter
depend on $tan\beta \equiv {{<{H_2}>} \over {<H_1>}} $,
which in turn depends on the details of  $SU(2)_L\times U(1)_Y$ breaking.
However, as mentioned above, the inclusion of D-terms leads
only to small shifts. In particular, for the extreme case of maximum D-term
contributions (large  $tan\beta$), the only modification is that the
$e_L$ and $e_L^c$ masses get slightly shifted upwards (a few GeV)
whereas the $v_L$ mass gets lowered (even below  $e_L^c$,
if the gluino is lighter than 350 GeV). The contribution to the scalar mass
RGEs of the $U(1)_Y$ D-term are vanishing due to their universal boundary
conditions.

The rest of the supersymmetric mass spectra are  more dependent on the
$SU(2)_L\times U(1)_Y$
breaking process and the value assumed for the $B$ parameter.
(The A-parameter which appears through the RGEs is given by
eq.(\ref{F1004}) with $\omega _{ijk}=0$ in the large-$T$ limit of Calabi--Yau
compactifications as explained in subsection 3.3.2. In the orbifold case is
given by eq.(\ref{F44}) with the last term in $\omega _{ijk}$ vanishing as
explained
in subsection 3.3.1.)
Indeed, these
scenarios are five-parameter
models in general: $m_{3/2}$, $\sin\theta$ (recall that one is restricted to a
region fulfilling eq.(\ref{F03})), $B$, the top-quark mass
$m_t$  and the $\hat{\mu}$ parameter.
One can eliminate one of them (e.g. $\hat{\mu}$,
 which is the one we know the least) in terms of the others by imposing
appropriate
symmetry breaking at the weak scale. The value of the top mass is quite
constrained by
the LEP and CDF data so that $m_t$ is not a source of big uncertainty.
However, $B$ introduces some source of uncertainty.
In general, $B$ is a model dependent function of $m_{3/2}, \sin\theta, T$,
etc.,
as discussed
in subsection 3.1. One could leave $B$ as a free parameter in the analysis,
but we find
more interesting to display the results in terms of some reasonable ansatz for
the
$\sin\theta $ dependence of the $B$ parameter.
One possibility is to use $B_{\mu}$ (eq.(\ref{F444}))
with the simplifying assumption $\mu^S/\mu=\mu^T/\mu=0$. This analysis can be
found in section 7
of ref\cite{BIM}. But we think it is still more interesting to use any of the
two mechanisms to generate the $\mu$ term discussed in section 2 and 3.1.
Now $\mu$ is no longer a free parameter and $B_{\lambda,Z}$ are completely
determined (see eqs.(\ref{F445},\ref{F45}). The analysis
using $B_Z$ can be found in ref.\cite{BIMS}.

Thus {\it we have traded} the four free soft parameters ($M,m,A,B$) of the MSSM
by the two parameters $m_{3/2}$ and $\theta $.

\vspace{0.2cm}

\noindent {\it The $sin\theta\rightarrow 0$ (modulus-dominated) limit}

\vspace{0.2cm}

\noindent There is only
one situation in which {\it the gaugino masses may become
smaller than the scalar masses}: a very small $\sin\theta$.
We recall that this limit
corresponds to a modulus-dominated SUSY-breaking even if $S$-$T$ mixing is
present, as discussed above eq.(\ref{F50}). This limit
{\it is only accesible if}
all modular weights of sparticles are equal to $-1$ (see eq.(\ref{F03})), as
in the large-$T$ Calabi--Yau limit and possible orbifold Superstring
scenarios. In this
case, $\sin\theta\rightarrow 0$ implies $M_a, m_i\rightarrow 0$ as can be
obtained from eqs.(\ref{F01},\ref{F02}) and therefore,
as $\sin\theta$ decreases,
one can ignore less and less the one-loop corrections to the K\"ahler
potential. Furthermore,
the results get more and more dependent on the form of the Superstring
threshold
correction
function $f_a(T)$ (see eq.(\ref{F26})),
which is still quite uncertain in the context of
Calabi--Yau-type
compactifications. On the other hand, as we studied already in
subsection 3.3.1,
both the
one-loop threshold effects and the one-loop corrections to the K\"ahler
potential
are much better known in the context of orbifold four-dimensional
Superstrings. Thus it makes
sense
to study the orbifold analogous to the Calabi--Yau
scenario, which will explicitely provide us
with one-loop corrected expressions for the soft terms.

Let us now describe the form of the soft masses in this scenario, starting
with the scalar masses. They can be obtained from eq.(\ref{F02}) with
$n_i=-1$
after introducing the one-loop correction eq.(\ref{F49}).
One gets
\begin{eqnarray}
m_i^2\ =\ m_{3/2}^2\ \left[1\ -\ \left(1-{{\delta _{GS}}\over {24\pi ^2
Y}}\right)^{-1}
cos^2\theta \right] \ ,
\label{F001}
\end{eqnarray}
where $Y$ was defined in eq.(\ref{F46}).
This result is numerically very similar to that studied above with $n_i=-1$
as long as $\sin\theta $ is not much smaller than one.
However, in the $\sin\theta \rightarrow 0 $ limit in which the tree-level
scalar masses vanish one finds
\begin{eqnarray}
m_i^2\ (sin\theta \rightarrow 0)\ \simeq \
m_{3/2}^2\ ({{-\delta _{GS}}\over {24\pi^2 Y}})\ \simeq\ m_{3/2}^2 (-\delta
_{GS})
10^{-3} \ .
\label{F002}
\end{eqnarray}
We thus observe that, in the case of orbifolds, the inclusion of the
one-loop corrections in the K\"ahler potential has the effect of "regulating"
in some
way the $\sin\theta \rightarrow 0$ limit yielding a non-vanishing result
for the scalar masses.

Finally, concerning the gaugino masses,
they can be obtained from eq.(\ref{F42}), with $C=1$ and vanishing phases,
after introducing the one-loop
correction from eqs.(\ref{F46},\ref{F49}). One gets
\begin{eqnarray}
M_a &=& {{k_a Y}\over {2 Ref_a}} {\sqrt 3}m_{3/2}\left[sin\theta\right.
\nonumber\\
&+& \left. {{(b_a'-k_a\delta _{GS})(T+T^*){\hat {G_2}}(T,T^*)}\over
{ 16{\sqrt 3}\pi^3 k_a Y}}
\left(1-{{\delta _{GS}}\over {24\pi ^2 Y}}\right)^{-1/2}
cos\theta \right] \ .
\label{F003}
\end{eqnarray}
%
%
Since all modular weights are equal to $-1$, one has $b_a'=b_a$
(see eq.(\ref{F41})). Then, using a value $ReT\simeq 1.2$
(this value is suggested by several gaugino condensation analyses,
see e.g. ref.\cite{Beatriz} and references
therein) close to the
duality self-dual point, which is what one would normally expect in a duality
invariant theory, one obtains the following
numerical results:
\begin{eqnarray}
M_3 \ & \simeq & \ 1.0\ {\sqrt 3}m_{3/2}\left[ sin\theta \ -\ (3+{\delta
_{GS}})\ 4.6\times 10^{-4}\ cos\theta \right] \ ,
\nonumber\\
M_2 \ & \simeq & \ 1.06 \ {\sqrt 3} m_{3/2} \left[ sin\theta \ -
(-1+{\delta_{GS}})\ 4.6\times 10^{-4}\ cos\theta \right] \ ,
\nonumber\\
M_1 \ &  \simeq & \ 1.18 \ {\sqrt 3}m_{3/2} \left[ sin\theta \ - ({-33\over
5}+{\delta _{GS}})\  4.6\times 10^{-4} \ cos\theta \right] \ .
\label{F004}
\end{eqnarray}
We are really interested in
understanding the qualitative behaviour of this small
$\sin\theta$ limit and hence we will just take a fixed
value for $\delta _{GS}$, e.g. $\delta _{GS}=-5$. This is a negative integer
with a
magnitude of order of the $b'$ coefficients involved and hence it is not an
unreasonable
value. We will comment below what happens as we vary this parameter.
The most prominent feature of the soft
terms is that {\it for values of $|sin\theta |$ below $5\times 10^{-2}$ the
gaugino masses
become smaller than the scalar masses}.
\begin{eqnarray}
M_a < m_i \ .
\label{F0440}
\end{eqnarray}
This is something which is
qualitatively
different from the previous results, where gaugino masses were
always necessarily larger than scalar masses.  On the other side,
this situation is quite similar to the one obtained in explicit
gaugino condensation models\cite{Beatriz2} although
it is not really identical (for a comparison of both situations, see section 8
of ref.\cite{BIM}).
Indeed, in this scenario
as $\sin\theta$ decreases the gaugino/squark mass ratio decreases.

Notice that the qualitative behaviour found here for small $\sin\theta $ is
generic
for any non-vanishing negative integer $\delta _{GS}$. The only difference
is the particular value of $\sin\theta $ at which the gaugino masses start
being
smaller than the scalar masses. Also, different values for $\delta _{GS}$ lead
to different gaugino mass ratios (e.g. $M_3/M_2$) as $\sin\theta \rightarrow
0$,
but
we consider this as a small correction to the most relevant feature found in
this
limit, which is that the gaugino masses become small compared to the scalar
masses. (The case
$\delta _{GS}=0$ is special since, as can be seen from eq.(\ref{F002}),
the scalar masses tend to
zero and the gaugino masses provide essentially the only source of
SUSY-breaking for
$\sin\theta =0$. $\delta _{GS}=0$ is e.g. the case of the orbifold
$Z_2 \times Z_2$, however, generically there will be $S$-$T$ mixing
in the K\"ahler potential and a case with $\delta _{GS}=0$ is atypical.)
Another point to remark is that in the present limit
the gravitino mass is much larger (more than an order of magnitude bigger) than
the soft masses. For example, for $\delta _{GS}=-5$, eq.(\ref{F002})
implies that $m_{3/2} \simeq 14 m_i$.

Let us describe now what is the structure of the low-energy SUSY-spectra in
this
small $\sin\theta $ limit. Let us start as usual with
the sector of the spectrum which is rather insensitive to the radiative
electroweak breaking, i.e. the gluino, the
squarks (except stops and left sbottom) and the sleptons.
Since the soft terms
have a different dependence on $\theta $, we will content ourselves with
showing results for a fixed small value of $\sin\theta $, because this is the
limit we want to explore (for large $\sin\theta $ the results correspond to
a good approximation with those studied above for $n_i=-1$).
For the illustrative choice
$\theta-\pi=5\times10^{-3}$,
with $ReT\simeq 1.2$ and $\delta _{GS}=-5$,
the situation now is completely reversed with respect to the above one.
The gluino is substantially lighter
than the scalars.
\begin{eqnarray}
M_g < m_l \simeq m_q .
\label{F0990}
\end{eqnarray}
For example,
in this particular case the relation is $m_{q,l} \simeq
2.5 M_g$. Notice also that the physical masses of squarks and sleptons
are almost degenerate. This happens because the universality
of soft scalar masses at high energy is not destroyed by the
gluino contribution to the mass renormalization, which is now very small.

The above results tell us that {\it if gluinos lighter than squarks and
sleptons are found,
this could be an indication that the dominant source of SUSY-breaking lies in
the moduli} and not in the dilaton sector. Concerning the rest of the
spectrum, similar remarks as above apply.

\vspace{0.2cm}

\noindent {\it Discussion of the overall Supersymmetric-spectra}

\vspace{0.2cm}

\noindent Let us try to
summarize the most prominent patterns obtained for the spectra
of supersymmetric particles in this large class of models. For a given
choice of Superstring model the free soft
parameters of the MSSM ($M,m,A,B$) are {\it given in terms of the
gravitino mass $m_{3/2}$ and the goldstino angle $\theta $}. In some
Superstring models in which the one-loop corrections become important
additional dependence on other parameters ($\delta _{GS},
ReT$) may appear, although the latter are less crucial in understanding the
{\it qualitative} patterns of soft terms.

One first point to remark is that one can have flavour-independent
soft scalar masses even without dilaton-dominated SUSY-breaking. In fact, all
the scenarios discussed above have flavour-independent scalar masses, although
in some case different scalars within the same
flavour generation can have different masses. Thus {\it dilaton-dominated
SUSY-breaking is a sufficient but not necessary condition to obtain
scalar mass universality}.

For goldstino angle $|\sin\theta|\geq 5\times 10^{-2}$ the results for the
different scenarios are not terribly different. The heaviest particles are
the coloured ones with gluinos and squarks almost degenerate and $3$ to $6$
times heavier than sleptons.

For very small $\sin\theta$ one can no longer neglect in general the one-loop
corrections. Now the situation concerning the spectrum is very much changed
and the gluino may be even lighter than the squarks and sleptons. The latter
become almost degenerate. If a spectrum of this type
is found, it could be an indication of a modulus-dominated SUSY-breaking.

It is also important to recall that values $\sin^2\theta \le 1/2$ are
only possible in models in which all modular weights are equal to $-1$.
Otherwise some of the squarks and/or sleptons would get negative
squared masses at the Superstring scale.

\vspace{0.2cm}

\noindent {\it The case with several moduli}

\vspace{0.2cm}

\noindent The formulae
for soft terms in the case of several moduli were written at
the end of subsection 3.3.1 (see eq.(\ref{F97})). Taking into account these
results is possible to see that the general phenomenological conclusions
obtained above for the overall modulus case may be somewhat
modified\cite{BIMS}.
For example, particles with $n_i=-1$ may also get negative mass-squared
for some choices of the angles. Also, scalar masses may become bigger than
gaugino masses even at tree level. For an extended discussion see
ref.\cite{BIMS}.

\vspace{0.2cm}

\noindent {\it Final comment about the spectra}

\vspace{0.2cm}

The general pattern of SUSY-spectra found in the present approach
are very characteristic.
 Optimistically, if the spectrum of
SUSY particles is eventually found, one will be able to rule out
(or rule in) some of the general scenarios (e.g., dilaton or
modulus dominance) here discussed. More modestly, we hope that
the formulae and the examples worked out in this paper will be
of help in looking for a more fundamental understanding of the
origin of SUSY-breaking soft terms in the Supersymmetric Standard
Model.

\section{Acknowledgements}

We thank A. Lleyda for her valuable help.


\section{References}

\begin{thebibliography}{9}

\bibitem{Rusos} Yu.A. Gol'fand and E.P. Likhtman,
{\it JETP Lett.} {\bf 13} (1971) 323;
D.V. Volkov and V.P. Akulov, {\it JETP Lett.} {\bf 16} (1972) 438;
J. Wess and B. Zumino,
{\it Nucl. Phys.} {\bf B70} (1974) 39.
\bibitem{Nilles} For a review, see: H.P. Nilles, {\it Phys.
Rep.} {\bf 110} (1984) 1, and references therein.
\bibitem{Zumino} D.Z. Freedman, P. Van Nieuwenhuizen and S. Ferrara,
{\it Phys. Rev.} {\bf D13} (1976) 3214; S. Deser and B. Zumino,
{\it Phys. Lett.} {\bf 62B} (1976) 335.
\bibitem{history} For a historical review, see: S. Ferrara, {\it
Dirac Lecture} delivered at ICTP, Trieste (1994), CERN-TH.7285/94,
hep-th/9405065, and references therein.
\bibitem{SW} J. Scherk and J.H. Schwarz,
{\it Nucl. Phys.} {\bf B81} (1974) 118; M.B. Green and J.H. Schwarz,
{\it Phys. Lett.} {\bf 149B} (1984) 117.
\bibitem{Green} For a review, see: M. Green, J.H. Schwarz and E. Witten,
{\it Superstring Theory} (Cambridge University Press, 1987), and references
therein.
\bibitem{history2} For a historical review, see: J.H. Schwarz, {\it
Dirac Lecture} delivered at ICTP, Trieste (1989), and references therein.
\bibitem{Grisaru} L. Girardello and M.T. Grisaru,
{\it Nucl. Phys.} {\bf B194} (1982) 65.
\bibitem{Hall} L. Hall, J. Lykken and S. Weinberg, {\it Phys. Rev.} {\bf D27}
(1983) 2359.
\bibitem{Soni} S.K. Soni and H.A. Weldon,
{\it Phys. Lett.} {\bf B126} (1983) 215.

\bibitem{GC} H.P. Nilles, Phys. Lett. B115 (1982) 193;
{\it Nucl. Phys.} {\bf B217} (1983) 366; S. Ferrara,
L. Girardello and H.P. Nilles,
{\it Phys. Lett.} {\bf B125} (1983) 457;
J.P. Derendinger, L.E. Ib\'{a}\~{n}ez and
H.P. Nilles,
{\it Phys. Lett.} {\bf B155} (1985) 65;
M. Dine, R. Rohm, N. Seiberg and E. Witten,
{\it Phys. Lett.} {\bf 156B} (1985) 55.

\bibitem{Nilles2} For a review, see: H.P. Nilles, {\it Int. J. Mod. Phys.}
{\bf A5} (1990) 4199, and references therein.
\bibitem{Sugra} E. Cremmer, S. Ferrara, L. Girardello and A. Van
Proeyen, {\it Nucl. Phys.} {\bf B212} (1983) 413.
\bibitem{Modelos} B. Greene, K.H. Kirklin, P.J. Miron and G.G. Ross,
{\it Phys. Lett.} {\bf B180} (1986) 69;
J.A. Casas, E.K. Katehou and C. Mu\~{n}oz,
{\it Nucl. Phys.} {\bf B317} (1989) 171;
J.A. Casas and C. Mu\~{n}oz,
{\it Phys. Lett.} {\bf B214} (1988) 63;
A. Font, L. Ib\'{a}\~{n}ez, H.P. Nilles and F. Quevedo,
{\it Phys. Lett.} {\bf B210} (1988) 101;
I. Antoniadis, J. Ellis, J.S. Hagelin and D.V. Nanopoulos,
{\it Phys. Lett.} {\bf B205} (1988) 459.
\bibitem{Search} For a review, see e.g.: J.A. Casas and C. Mu\~{n}oz,
{\it Nucl. Phys. (Proc. Suppl.)} {\bf B16} (1990) 624, and references therein.
\bibitem{Witten} E. Witten, {\it Phys. Lett.} {\bf 155B} (1985) 151.
\bibitem{Kaplunovsky} V.S. Kaplunovsky,
{\it Nucl. Phys.} {\bf B307} (1988) 145 [Erratum: {\bf B382} (1992) 436].
\bibitem{Yo} For a brief review, see e.g.: C. Mu\~noz, {\it talk} given at the
International Conference "Beyond the Standard Model IV", Lake Tahoe,
(California), 1994, FTUAM 95/5, hep-ph/9503314, and references therein.
\bibitem{Hamidi} S. Hamidi and C. Vafa, {\it Nucl. Phys.} {\bf B279} (1987)
465; L. Dixon, D. Friedan, E. Martinec and S. Shenker, {\it Nucl. Phys.}
{\bf B282} (1987) 13.
\bibitem{Faustino} See e.g.: J.A. Casas, F. G\'{o}mez and C. Mu\~noz,
{\it Phys. Lett.} {\bf B292} (1992) 42, and references therein.
\bibitem{Gross} D.J. Gross, J.A. Harvey, E. Martinec and R. Rohm,
{\it Nucl. Phys.} {\bf B256} (1985) 256; {\bf B267} (1986) 75.
\bibitem{FKP} S. Ferrara, C. Kounnas and M. Porrati,
{\it Phys. Lett.} {\bf B181} (1986) 263; M. Cveti\u{c},
J. Louis and B. Ovrut, {\it Phys. Lett.} {\bf B206} (1988) 227;
M. Cveti\u{c}, J. Molera and B. Ovrut, {\it Phys. Rev.} {\bf D40} (1989) 1140.
\bibitem{DKL1} L.J. Dixon, V.S. Kaplunovsky and J. Louis,
{\it Nucl. Phys.} {\bf B329} (1990) 27.
\bibitem{DKL2} L.J. Dixon, V.S. Kaplunovsky and J. Louis,
{\it Nucl. Phys.} {\bf B355} (1991) 649.
\bibitem{Louis} J. Louis, {\it Proceedings} Boston PASCOS (1991) 751.
\bibitem{BIM} A. Brignole, L.E. Ib\'{a}\~{n}ez and C. Mu\~noz,
{\it Nucl. Phys.} {\bf B422} (1994) 125 [Erratum: {\bf B436} (1995) 747].
\bibitem{Beatriz} B. de Carlos, J.A. Casas and C. Mu\~{n}oz,
{\it Nucl. Phys.} {\bf B399} (1993) 623.
\bibitem{Beatriz2} B. de Carlos, J.A. Casas and C. Mu\~{n}oz,
{\it Phys. Lett.} {\bf B299} (1993) 234.
\bibitem{cc} S. Weinberg, {\it Rev. Mod. Phys.} {\bf 61} (1989) 1.
\bibitem{Gaugino} A. Font, L.E. Ib\'{a}\~{n}ez, D. L\"{u}st and
F. Quevedo, {\it Phys. Lett.} {\bf B245} (1990) 401.
\bibitem{Cvetic} M. Cveti\u{c}, A. Font, L.E. Ib\'{a}\~{n}ez, D. L\"{u}st and
F. Quevedo, {\it Nucl. Phys.} {\bf B361} (1991) 194.
\bibitem{IL} L.E. Ib\'{a}\~{n}ez and D. L\"ust,
{\it Nucl. Phys.} {\bf B382} (1992) 305.
\bibitem{KapluLouis} V.S. Kaplunovsky and J. Louis
{\it Phys. Lett.} {\bf B306} (1993) 269.
\bibitem{Ferrara} S. Ferrara, C. Kounnas and F. Zwirner,
{\it Nucl. Phys.} {\bf B429} (1994) 589 [Erratum: {\bf B433} (1995) 255].
\bibitem{Horowitz} P. Candelas, G. Horowitz, A. Strominger and E. Witten,
{\it Nucl. Phys.} {\bf B258} (1985) 46.
\bibitem{Harvey} L.J. Dixon, J. Harvey, C. Vafa and E. Witten,
{\it Nucl. Phys.} {\bf B261} (1985) 651; {\bf B274} (1986) 285.
\bibitem{Bfs} R. Barbieri, S. Ferrara and C.A. Savoy,
{\it Phys. Lett.} {\bf 119B} (1982) 343.
\bibitem{Rossbook} G.G. Ross, {\it Grand Unified Theories}
(Benjamin/Cummings Publishing Co., 1984).
\bibitem{Peccei} R. Peccei and H. Quinn, {\it Phys. Rev. Lett.} {\bf 38}
(1977) 1440.
\bibitem{Weinberg} S. Weinberg, {\it Phys. Rev. Lett.} {\bf 40} (1978) 223;
F. Wilczeck, {\it Phys. Rev. Lett.} {\bf 40} (1978) 229.
\bibitem{Kim} J.E. Kim and H.P. Nilles, {\it Phys. Lett.} {\bf B138}
(1984) 150, {\bf B263} (1991) 79;
E.J. Chun, J.E. Kim and H.P. Nilles, {\it Nucl. Phys.} {\bf B370} (1992) 105.
\bibitem{Masiero} G.F. Giudice and A. Masiero, {\it Phys. Lett.} {\bf B206}
(1988) 480.
\bibitem{Casas} J.A. Casas and C. Mu\~noz, {\it Phys. Lett.} {\bf B306}
(1993) 288.
\bibitem{Giudice} G.F. Giudice and E. Roulet, {\it Phys. Lett.} {\bf B315}
(1993) 107.
\bibitem{Ignatios} I. Antoniadis, C. Mu\~noz and M. Quir\'{o}s
{\it Nucl. Phys.} {\bf B397} (1993) 515;
S. Ferrara, C. Kounnas, M. Porrati, F. Zwirner {\it Nucl. Phys.} {\bf B318}
(1989) 75.
\bibitem{Narain} I. Antoniadis, E. Gava, K.S. Narain and T.R. Taylor,
{\it Nucl. Phys.} {\bf B432} (1994) 187.
\bibitem{Dieter} G. Lopes-Cardoso, D. L\"ust and T. Mohaupt,
{\it Nucl. Phys.} {\bf B432} (1994) 68.
\bibitem{IN} L.E. Ib\'{a}\~{n}ez and H.P. Nilles, {\it Phys. Lett.} {\bf 169B}
(1986) 354.
\bibitem{higher} H.P. Nilles, {\it Phys. Lett.} {\bf 180B} (1986) 240;
M.A. Shifman and A.I. Vainshtein, {\it Nucl. Phys.} {\bf B359}
(1991) 571; J.A. Casas and C. Mu\~noz, {\it Phys. Lett.} {\bf B271}
(1991) 85; I. Antoniadis, K.S. Narain and T.R. Taylor,
{\it Phys. Lett.} {\bf B267} (1991) 37.
\bibitem{BIMS} A. Brignole,
L.E. Ib\'{a}\~{n}ez, C. Mu\~noz and C. Scheich, to appear.
\bibitem{Pokorski} D. Choudhury, F. Eberlein, A. K\"oning, J. Louis and
S. Pokorski, MPI-Ph/94-51, hep-ph/9408275.
\bibitem{Nir} J. Louis and Y. Nir, LMU-TPW 94-17, hep-ph/9411429.
\bibitem{Choi} K. Choi, {\it Phys. Rev. Lett.} {\bf 72} (1994) 1592.
\bibitem{KN} K. Choi, J.E. Kim and H.P. Nilles,
{\it Phys. Rev. Lett.} {\bf 73} (1994) 1758.
\bibitem{Park} K. Choi, J.E. Kim and G.T. Park, SNUTP 94-94, hep-ph/9412397.
\bibitem{Barbieri} R. Barbieri, J. Louis and M. Moretti,
{\it Phys. Lett.} {\bf B312} (1993) 451.
\bibitem{FLT} S. Ferrara, D. L\"{u}st A. Shapere and S. Theisen,
{\it Phys. Lett.} {\bf B225} (1989) 363;
S. Ferrara, D. L\"{u}st and S. Theisen, Phys. Lett.
{\it Phys. Lett.} {\bf B233} (1989) 147.
\bibitem{DFKZ} J.P. Derendinger, S. Ferrara, C. Kounnas and F. Zwirner,
{\it Nucl. Phys.} {\bf B372} (1992) 145; {\it Phys. Lett.} {\bf B271} (1991)
307.
\bibitem{Cual} I. Antoniadis, E. Gava and K.S. Narain,
{\it Nucl. Phys.} {\bf B383} (1992) 93.
\bibitem{Stieberger} P. Mayr and S. Stieberger, {\it Nucl. Phys.} {\bf B407}
(1993) 425; {\it Nucl. Phys.} {\bf B412} (1994) 502.
\bibitem{Bailin} D. Bailin, A. Love, W.A. Sabra and S. Thomas,
{\it Mod. Phys. Lett.} {\bf A9} (1994) 67.
\bibitem{Schwarz} M.B. Green and J.H. Schwarz, {\it Phys. Lett.} {\bf 149B}
(1984) 117.
\bibitem{Ovrut} G. Lopes Cardoso and B. Ovrut, {\it Nucl. Phys.} {\bf B369}
(1992) 351.
\bibitem{Unpub} A. Brignole, L.E. Ib\'{a}\~{n}ez and C. Mu\~noz,
unpublished (1993).
\bibitem{Japoneses} T. Kobayashi, D. Suematsu, K. Yamada and Y. Yamagishi,
{\it Phys. Lett.} {\bf B348} (1995) 402.
\bibitem{Candelas} P. Candelas, X.C. de la Ossa, P.S. Green and L. Parkes,
{\it Phys. Lett.} {\bf B258} (1991) 118; {\it Nucl. Phys.} {\bf B359} (1991)
21.
\bibitem{Dixonunpub} L.J. Dixon, unpublished.
\bibitem{Lustcvetic} S. Ferrara, C. Kounnas, D. L\"{u}st and F. Zwirner,
{\it Nucl. Phys.} {\bf B365} (1991) 431;
M. Cveti\u{c}, talk given at the Int. Conf. on High energy physics, Dallas,
1992.
\bibitem{ILR} L.E. Ib\'{a}\~{n}ez, D. L\"{u}st and G.G. Ross,
{\it Phys. Lett.} {\bf B272} (1991) 251.
\bibitem{EllisCremmer} For a review, see: A.B. Lahanas and D.V. Nanopoulos
{\it Phys. Rep.} {\bf 145} (1987) 1, and references therein.
\bibitem{Amanda} A. Lleyda and C. Mu\~{n}oz,
{\it Phys. Lett.} {\bf B317} (1993) 82.


\end{thebibliography}
\end{document}